\documentclass[10pt,twoside]{article}
\usepackage{latexsym}
\usepackage{graphicx}
\usepackage[english]{babel}
\usepackage{amssymb,amsmath,amsfonts,amsthm}
\usepackage{mathrsfs}
\usepackage{fancyhdr}
\usepackage{feynmp}

\newcommand{\be}{\begin{equation}}
\newcommand{\ee}{\end{equation}}
\newcommand{\bea}{\begin{eqnarray}}
\newcommand{\eea}{\end{eqnarray}}
\newcommand{\beas}{\begin{eqnarray*}}
\newcommand{\eeas}{\end{eqnarray*}}
\newcommand{\ovr}{\overline}
\newcommand{\ba}{\begin{array}}
\newcommand{\ea}{\end{array}}

\def\eps{\epsilon}

\def\s{\sigma}

\def\a{\alpha}
\def\da{\dot\alpha}
\def\r{\rho}
\allowdisplaybreaks[1]

\begin{document}

\begin{fmffile}{dia}
\fmfset{arrow_len}{1.5mm} \fmfset{wiggly_len}{2mm} \fmfset{dot_size}{3} \fmfset{dot_len}{1mm}

\thispagestyle{empty}

\begin{flushright}
ROM2F/2005/05 \\
\end{flushright}

\vspace{1.5cm}

\begin{center}

{\LARGE {\bf A wide class of four point functions \\of BPS
  operators in $\mathcal N=4$ SYM at order $g^4$ \rule{0pt}{25pt} }} \\
\vspace{1cm} \ {Marco D'Alessandro and Luigi Genovese} \\
\vspace{0.6cm} { {\it Dipartimento di Fisica, \ Universit\`a di Roma \ ``Tor Vergata''}} \\  {{\it I.N.F.N.\ -\ Sezione
di Roma \ ``Tor Vergata''}} \\ {{\it Via della Ricerca  Scientifica, 1}}
\\ {{\it 00133 \ Roma, \ ITALY}}
\end{center}

\vspace{1cm}

\begin{abstract}
The calculation of a large family
of four point functions of general BPS operators in $\mathcal N=4$ SYM
is reduced to the evaluation of colour contractions.
For $\frac{1}{2}$ BPS operators $\mathcal O_{\Delta}$ the explicit
results at order $g^4$ for the function
$\langle \mathcal O_n \mathcal O_2 \mathcal O_n \mathcal O_2 \rangle$ are given up to $n=6$.
The OPE of the general result is performed up to the second order in
the short distance expansion parameter.
Two examples are given, in which the mixing of the operators in the intermediate channel can
be resolved using four point functions computed by this method.
\end{abstract}
\newpage

\setcounter{page}{1}

\section{Introduction}
The pedagogical usefulness of $\mathcal N = 4$ SYM has been widely
evidenced in the recent years. This theory has been extensively studied both for its relevance to the AdS/CFT framework
(for reviews see \cite{Aharony:1999ti})
and for its purely QFT aspects, in which it can be viewed as the simplest interacting (super)conformal field theory
(CFT), completely determined by the choice of the gauge group and known to be finite
\cite{Grisaru:1980nk}. In the superconformal phase the spectrum of this
theory is very rich, and it is arranged in representations of the supergroup $SU(2,2|4)$, that contains the four
dimensional conformal group $SO(4,2)$ and the R-symmetry group $SU(4)$. The operators of the theory are thus organized
in superconformal multiplets, in which there are both primary and descendant conformal fields. In each supermultiplet,
the conformal primaries can be identified with their scaling dimension
$\Delta = \Delta^{(0)} + \gamma(g^2)$, the spin $(j_1,j_2)$ and the Dynkin
labels $[p,q,r]$ of the R-symmetry representation to which they
belong.
The unique quantum number that depends of the coupling constant
$g$ is the anomalous dimension $\gamma(g^2)$, that is the same for the
whole supermultiplet.
It is useful to think of
$\mathcal N = 4$ SYM as a CFT with additional symmetries that further constrain the dynamics. From the $\mathcal N = 4$
SYM calculations we can infer general properties of ordinary CFT's.

In a general four dimensional CFT, conformal invariance puts tight constraints on the correlation functions of the
primary operators \cite{Ferrara:1974pt}. Two and three point functions are completely fixed, up to
multiplicative constants related to the trilinear couplings of the theory\footnote{The ratio of the square of the three
point function coefficients and the product of the ones relative to the two point functions does not depend on the
chosen operator normalization and can be interpreted as the invariant trilinear coupling.}.
The freedom in the four point function are arbitrary functions of the conformal invariant cross ratios.
In a CFT, these correlators are thus the first ones whose space-time dependence is not fixed by the symmetries.
In a theory in which Operator Product Expansion (OPE) holds,
one can relate the knowledge of these correlators to the anomalous dimensions and the OPE structure constants,
related to the trilinear couplings of the operators in the theory via the two point function normalization.
The presence
of an operator in an OPE channel is dictated by the fusion rules,
that determine the conformal families of operators which can appear in the product of the two fields.
Since the OPE implements conformal invariance, if one can obtain all the OPE structure constants and
the anomalous dimensions of the operators of the theory,
in principle all the $n$-point functions can be calculated.

For these reasons the four point functions of operators in $\mathcal N = 4$ SYM have been extensively studied
\cite{Eden:1998hh,Eden:1999kh,Bianchi:1999ge,Bianchi:2000hn,Bianchi:2001cm,Arutyunov:2001mh,Arutyunov:2003ad,Arutyunov:2002fh,Dolan:2004iy}.
Particular attention was given to correlators involving protected
operators that obey shortening conditions, and belong to short $SU(2,2|4)$ representations \cite{Dobrev:1985qv,Andrianopoli:1999vr,Eden:2001ec}.
Correlation functions of such operators have additional non renormalization properties
\cite{Lee:1998bx,Eden:1999kw,nonrenorm},
that can be used to simplify the calculations.

These functions exhibit short distance logarithmic singularities,
whose coefficients can be interpreted in terms of the anomalous dimensions and trilinear couplings of the operators in the theory.
Although the operators in the external points are protected,
to their four point correlators contribute also non protected fields,
that appear in the intermediate OPE channel.
Hence, the study of four point functions of operators belonging to short multiplets can give us a lot of far from trivial information,
also about non protected operators.

In this paper we will show how to reduce the calculation at order $g^4$ of a wide class of four point functions of protected operators
in a particular flavour projection, to the computation of some $SU(N)$ color contractions.
The described method is particularly efficient for four point functions of $\frac{1}{2}$ BPS operators,
due to the non renormalization properties of their (extremal) three point functions.
However, we will see that it can be applied also to other operators belonging to BPS multiplets.

We perform also the OPE analisys, up to the second order in the short distance expansion parameter, of this class of correlators.
If one is able to find a sufficient number of correlators in which the OPE intermediate channel contains the same operators,
it is possible to extract the anomalous dimensions
(up to order $g^4$) of the operators,
together with the  tree level and one loop  trilinear couplings of these operators with the external ones.

The paper is organized as follows: in section \ref{sec:funzioni a quattro punti protette} we present our method,
and we discuss under which conditions it can be applied also to non $\frac{1}{2}$ BPS operators.
In section \ref{sec:general OPE expansions} we perform an OPE expansion of the general result,
and derive a system of equations for the $\mathcal O (g^2)$ trilinear couplings and the
$\mathcal O (g^4)$ anomalous dimensions of the operators in that channel.
Then, in section \ref{sec:examples}, two examples of such a computation are given,
in which there are enough correlators to extract the information for the operators involved.
In the last section we comment our result and we give some outlook on how to further pursue the work following this approach.
Computational details and other results are presented in the appendix.

\section{Four point function of protected operators} \label{sec:funzioni a quattro punti protette}
Considering four point functions of protected operators presents several advantages.
In fact, provided the finiteness of the theory,
these four point functions are \emph{finite}, and there are no subtleties in the calculation due to the renormalization of the operators in the
external points. The logarithmic singularities we obtain by performing the OPE expansion depend only upon the short distance expansion parameter.

In this section we will show how to calculate exactly, up to order $g^4$, by simply performing color contractions,
all the four point function of scalar operators of the type
\be \label{eqn:prima definizione}
\langle {\mathcal{O}_L}_{[0,n,0]}^{\Delta=n} {\mathcal{O}}_{[0,2,0]}^{\Delta=2}{\mathcal{O}_R}_{[0,n,0]}^{\Delta=n} {\mathcal{O}}_{[0,2,0]}^{\Delta=2}\rangle \;,
\ee
for a particular choice of the flavour representatives, in the $\mathcal N=1$ formulation (see appendix).
This is a four point function of four $\frac{1}{2}$ BPS operators,
two of them having dimension 2, and the other two with equal dimension $\Delta=n$.

Actually, we can go further, this method can be used also for calculate four point functions with other protected
operators, not necessarily $\frac{1}{2}$ BPS.

\subsection{General setting}
Let's first define how we choose the operators in the external points. As explained before, we consider four scalar $\frac{1}{2}$ BPS
operators of integer dimension $\Delta=n$, in the $[0,n,0]$ representation of the R-symmetry group $SU(4)$.
In particular we will use two $\Delta=2$ and two $\Delta=n$ operators. The $\Delta=2$ operator is the scalar operator in the $20'$ representation,
belonging to the ultrashort supercurrent multiplet.
Written in the $N=1$ formalism (see appendix), the two representatives we choose are
\begin{align}
\mathcal C_{22} =& \text{tr} \Bigl(\phi_2 \phi_2\Bigr) \;,\\
\mathcal C_{11}^\dagger =& \text{tr} \left(\phi_1^\dagger \phi_1^\dagger \right)\;.
\end{align}
In general, there are several $\frac{1}{2}$ BPS scalar operators in the $[0,n,0]$ representation.
For our calculation it is convenient to make the following choice of the representatives:
\begin{align}\label{eqn:representatives}
\mathcal O_L^{11;3} =& X^L_{a b c_1 \cdots\, c_{n-2}}\, \phi_1^a \phi_1^b \phi_3^{c_1} \cdots \phi_3^{c_{n-2}} \;, \\
{\mathcal O_R^\dagger}^{22;3} =& X^R_{a b c_1 \cdots\, c_{n-2}}\, {\phi_2^\dagger}^a {\phi_2^\dagger}^b {\phi_3^\dagger}^{c_1} \cdots {\phi_3^\dagger}^{c_{n-2}}\nonumber
\;,
\end{align}
where the two color tensors $X^L$ and $X^R$ describe the
color contractions which define the operators. In order to describe $\frac{1}{2}$ BPS operators,
they have to be totally symmetric, but not necessarily traceless.

We have all the elements to define the four point function under consideration.
\begin{equation}\label{eqn:definizione di Gn}
G_n(x_1,x_2,x_3,x_4) \equiv \langle \mathcal O_L^{11;3}(x_1) \, \mathcal C_{11}^\dagger(x_2) \, {\mathcal O_R^\dagger}^{22;3}(x_3) \, \mathcal C_{22}(x_4)
\rangle\;.
\end{equation}
Actually the function $G_n$ has also indices $L$, $R$ that we omit for simplicity.
This function can be expanded in powers of the coupling constant $g$, giving
\be \label{eqn:espansione in potenze di Gn}
G_n(x_1,x_2,x_3,x_4)=G_n^{(0)}(x_1,x_2,x_3,x_4)+g^2\,G_n^{(1)}(x_1,x_2,x_3,x_4)+g^4\,G_n^{(2)}(x_1,x_2,x_3,x_4)+\cdots\;,
\ee
and can be computed order by order in perturbation theory.
The computation passes through the calculation of several Feynman diagrams, that can be organized in different classes.
This classification will help us in developing an order $g^4$ computation.
From this result we can extract a lot of information,
by performing the OPE and analyzing the short distance logarithmic singularities which appear.
We will see in section \ref{sec:general OPE expansions} that for our purposes the non singular part in the
short distance expansion at order $g^4$ can be safely neglected, without lose of information.

\subsection{Classification of diagrams}
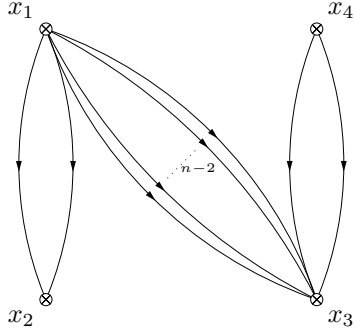
\begin{figure}[htbp]
\begin{center}
\begin{fmfgraph*}(45,45)
 \fmffreeze
  \fmfpen{.4}

    \fmfforce{0.1w,0.9h}{v1}
    \fmfforce{0.1w,0.9h}{v1a}
    \fmfforce{0.1w,0.1h}{v2}
    \fmfforce{0.1w,0.1h}{v2a}
    \fmfforce{0.9w,0.1h}{v3}
    \fmfforce{0.9w,0.1h}{v3a}
    \fmfforce{0.9w,0.9h}{v4}
    \fmfforce{0.9w,0.9h}{v4a}

    \fmfforce{0.46w,0.46h}{c1}
    \fmfforce{0.5w,0.5h}{c2}
    \fmfforce{0.54w,0.54h}{c3}

    \fmfv{d.sh=circle,d.fill=empty,d.si=2.3thick,l=$x_{1}$,l.a=135}{v1}
    \fmfv{d.sh=cross,d.si=2.3thick}{v1a}
    \fmfv{d.sh=circle,d.fill=empty,d.si=2.3thick,l=$x_{2}$,l.a=-135}{v2}
    \fmfv{d.sh=cross,d.si=2.3thick}{v2a}
    \fmfv{d.sh=circle,d.fill=empty,d.si=2.3thick,l=$x_{3}$,l.a=-45}{v3}
    \fmfv{d.sh=cross,d.si=2.3thick}{v3a}
    \fmfv{d.sh=circle,d.fill=empty,d.si=2.3thick,l=$x_{4}$,l.a=45}{v4}
    \fmfv{d.sh=cross,d.si=2.3thick}{v4a}

    \fmf{fermion,left=.20}{v1,v2}
    \fmf{fermion,right=.20}{v1,v2}
    \fmf{fermion,left=.20}{v4,v3}
    \fmf{fermion,right=.20}{v4,v3}
    \fmf{fermion,left=.23}{v1,v3}
    \fmf{fermion,right=.23}{v1,v3}
    \fmf{fermion,left=.16}{v1,v3}
    \fmf{fermion,right=.16}{v1,v3}
    \fmf{dots,label=$_{_{n-2}}$,label.dist=-.005w,label.side=right}{c1,c3}

\end{fmfgraph*}
\end{center}
\caption{Tree level diagram of $G_n$}\label{fig:treeleveldiagram}
\end{figure}

Let us now analyze the different contributions in the calculation of the correlator \eqref{eqn:definizione di Gn} at a generic pertubative order.
The tree-level diagram of this four point function is depicted in fig. \ref{fig:treeleveldiagram}.
Due to our choice of flavour representatives, there are three bundles of lines: from $x_1$ to $x_2$,
from $x_1$ to $x_3$ and from $x_4$ to $x_3$.

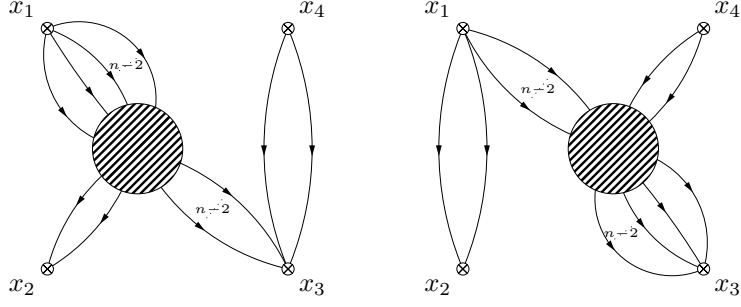
\begin{figure}[htbp]
\begin{center}
\begin{fmfgraph*}(40,40)
 \fmffreeze
  \fmfpen{.4}

    \fmfforce{0.1w,0.9h}{v1}
    \fmfforce{0.1w,0.9h}{v1a}
    \fmfforce{0.1w,0.1h}{v2}
    \fmfforce{0.1w,0.1h}{v2a}
    \fmfforce{0.9w,0.1h}{v3}
    \fmfforce{0.9w,0.1h}{v3a}
    \fmfforce{0.9w,0.9h}{v4}
    \fmfforce{0.9w,0.9h}{v4a}

    \fmfforce{0.63w,0.27h}{c1}
    \fmfforce{0.4w,0.5h}{c2}
    \fmfforce{0.68w,0.33h}{c3}
    \fmfforce{0.34w,0.76h}{c4}
    \fmfforce{0.395w,0.8h}{c5}

    \fmfv{d.sh=circle,d.fill=empty,d.si=2.3thick,l=$x_{1}$,l.a=135}{v1}
    \fmfv{d.sh=cross,d.si=2.3thick}{v1a}
    \fmfv{d.sh=circle,d.fill=empty,d.si=2.3thick,l=$x_{2}$,l.a=-135}{v2}
    \fmfv{d.sh=cross,d.si=2.3thick}{v2a}
    \fmfv{d.sh=circle,d.fill=empty,d.si=2.3thick,l=$x_{3}$,l.a=-45}{v3}
    \fmfv{d.sh=cross,d.si=2.3thick}{v3a}
    \fmfv{d.sh=circle,d.fill=empty,d.si=2.3thick,l=$x_{4}$,l.a=45}{v4}
    \fmfv{d.sh=cross,d.si=2.3thick}{v4a}

    \fmf{fermion,right=.05}{v1,c2}
    \fmf{fermion,right=.50}{v1,c2}
    \fmf{fermion,left=.20}{c2,v2}
    \fmf{fermion,right=.20}{c2,v2}
    \fmf{fermion,left=.20}{v4,v3}
    \fmf{fermion,right=.20}{v4,v3}
    \fmf{fermion,left=.3}{v1,c2}
    \fmf{fermion,left=.8}{v1,c2}
    \fmf{fermion,left=.23}{c2,v3}
    \fmf{fermion,right=.23}{c2,v3}
    \fmfblob{.3w}{c2}
    \fmf{dots,label=$_{_{_{n-2}}}$,label.dist=0w,label.side=right}{c1,c3}
    \fmf{dots,label=$_{_{_{n-2}}}$,label.dist=0w,label.side=left}{c4,c5}

\end{fmfgraph*}
\qquad\qquad
\begin{fmfgraph*}(40,40)
 \fmffreeze
  \fmfpen{.4}

    \fmfforce{0.1w,0.9h}{v1}
    \fmfforce{0.1w,0.9h}{v1a}
    \fmfforce{0.1w,0.1h}{v2}
    \fmfforce{0.1w,0.1h}{v2a}
    \fmfforce{0.9w,0.1h}{v3}
    \fmfforce{0.9w,0.1h}{v3a}
    \fmfforce{0.9w,0.9h}{v4}
    \fmfforce{0.9w,0.9h}{v4a}

    \fmfforce{0.33w,0.67h}{c1}
    \fmfforce{0.6w,0.5h}{c2}
    \fmfforce{0.38w,0.73h}{c3}
    \fmfforce{0.605w,0.2h}{c4}
    \fmfforce{0.66w,0.24h}{c5}

    \fmfv{d.sh=circle,d.fill=empty,d.si=2.3thick,l=$x_{1}$,l.a=135}{v1}
    \fmfv{d.sh=cross,d.si=2.3thick}{v1a}
    \fmfv{d.sh=circle,d.fill=empty,d.si=2.3thick,l=$x_{2}$,l.a=-135}{v2}
    \fmfv{d.sh=cross,d.si=2.3thick}{v2a}
    \fmfv{d.sh=circle,d.fill=empty,d.si=2.3thick,l=$x_{3}$,l.a=-45}{v3}
    \fmfv{d.sh=cross,d.si=2.3thick}{v3a}
    \fmfv{d.sh=circle,d.fill=empty,d.si=2.3thick,l=$x_{4}$,l.a=45}{v4}
    \fmfv{d.sh=cross,d.si=2.3thick}{v4a}

    \fmf{fermion,left=.05}{c2,v3}
    \fmf{fermion,left=.50}{c2,v3}
    \fmf{fermion,right=.20}{v4,c2}
    \fmf{fermion,left=.20}{v4,c2}
    \fmf{fermion,left=.20}{v1,v2}
    \fmf{fermion,right=.20}{v1,v2}
    \fmf{fermion,right=.3}{c2,v3}
    \fmf{fermion,right=.8}{c2,v3}
    \fmf{fermion,right=.23}{v1,c2}
    \fmf{fermion,left=.23}{v1,c2}
    \fmfblob{.3w}{c2}
    \fmf{dots,label=$_{_{_{n-2}}}$,label.dist=0w,label.side=left}{c1,c3}
    \fmf{dots,label=$_{_{_{n-2}}}$,label.dist=0w,label.side=right}{c4,c5}

\end{fmfgraph*}
\end{center}
\caption{Quantum correction reducible to three point functions}\label{fig:diagrammi funzioni a tre punti}
\end{figure}

The first type of quantum corrections that we analyze are the ones represented by the diagrams in figure
\ref{fig:diagrammi funzioni a tre punti}.
This means that we collect in this group all the interactions that do not involve the $x_4 - x_3$ or the $x_1-x_2$ bundle, respectively.
The core of the diagram, i.e. all the interaction lines and the other spectators that one can have, is represented here with the dark bubbles.
These quantum corrections vanish. As an example, let us consider the first class of diagrams.
Due to the presence of the spectator lines from $x_4$ to $x_3$ it is as if the operator in the point $x_3$ is
\be\label{eqn:definizione di Or'}
{\mathcal O^\dagger}^{3}_{R'} \equiv \delta^{ab} X^R_{a b c_1 \cdots\, c_{n-2}}\, {\phi_3^\dagger}^{c_1} \cdots {\phi_3^\dagger}^{c_{n-2}}\;.
\ee
This operator has bare dimension $n$ and belongs to the $[0,n,0]$ $SU(4)$ representation. So it is $\frac{1}{2}$ BPS.
Hence the calculation of these diagrams is equivalent to the computation of the correlation functions
\begin{equation}\label{eqn:funzioni a tre punti}
\langle \mathcal O_{L'}^{3} (x_1) \, {\mathcal O_R^\dagger}^{22;3}(x_3) \, \mathcal C_{22}(x_4) \rangle \;, \qquad \langle \mathcal O_L^{11;3}(x_1) \, \mathcal
C_{11}^\dagger(x_2) \, {\mathcal O^\dagger}^{3}_{R'}(x_3) \rangle \;,
\end{equation}
where $ {\mathcal O^\dagger}^{3}_{R'}$ is defined in \eqref{eqn:definizione di Or'} and the operator ${\mathcal O_{L'}^{3}}$ arises as a contraction of $\mathcal O_L^{11;3}$
with $\mathcal C_{11}^\dagger$:
\begin{align}
{\mathcal O_{L'}^{3} } &\equiv X^L_{a a c_1 \cdots\, c_{n-2}}\, \phi_3^{c_1} \cdots \phi_3^{c_{n-2}} \,, \label{eqn:definizione Ol'}\;.
\end{align}
The correlation functions indicated in \eqref{eqn:funzioni a tre punti} are (extremal) three   point functions of $\frac{1}{2}$
BPS operators, and therefore they do not receive quantum corrections \cite{Eden:1999kw,nonrenorm}.

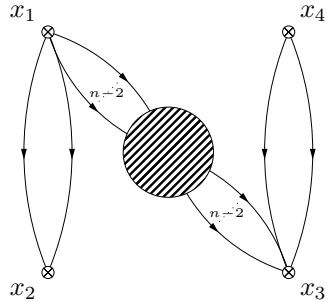
\begin{figure}[htbp]
\begin{center}
\begin{fmfgraph*}(40,40)
 \fmffreeze
  \fmfpen{.4}

    \fmfforce{0.1w,0.9h}{v1}
    \fmfforce{0.1w,0.9h}{v1a}
    \fmfforce{0.1w,0.1h}{v2}
    \fmfforce{0.1w,0.1h}{v2a}
    \fmfforce{0.9w,0.1h}{v3}
    \fmfforce{0.9w,0.1h}{v3a}
    \fmfforce{0.9w,0.9h}{v4}
    \fmfforce{0.9w,0.9h}{v4a}

    \fmfforce{0.27w,0.67h}{c1}
    \fmfforce{0.5w,0.5h}{c2}
    \fmfforce{0.33w,0.73h}{c3}
    \fmfforce{0.67w,0.27h}{c4}
    \fmfforce{0.73w,0.33h}{c5}

    \fmfv{d.sh=circle,d.fill=empty,d.si=2.3thick,l=$x_{1}$,l.a=135}{v1}
    \fmfv{d.sh=cross,d.si=2.3thick}{v1a}
    \fmfv{d.sh=circle,d.fill=empty,d.si=2.3thick,l=$x_{2}$,l.a=-135}{v2}
    \fmfv{d.sh=cross,d.si=2.3thick}{v2a}
    \fmfv{d.sh=circle,d.fill=empty,d.si=2.3thick,l=$x_{3}$,l.a=-45}{v3}
    \fmfv{d.sh=cross,d.si=2.3thick}{v3a}
    \fmfv{d.sh=circle,d.fill=empty,d.si=2.3thick,l=$x_{4}$,l.a=45}{v4}
    \fmfv{d.sh=cross,d.si=2.3thick}{v4a}

    \fmf{fermion,right=.25}{v1,c2}
    \fmf{fermion,left=.25}{v1,c2}
    \fmf{fermion,left=.20}{v1,v2}
    \fmf{fermion,right=.20}{v1,v2}
    \fmf{fermion,left=.20}{v4,v3}
    \fmf{fermion,right=.20}{v4,v3}
    \fmf{fermion,left=.25}{c2,v3}
    \fmf{fermion,right=.25}{c2,v3}
    \fmfblob{.3w}{c2}
    \fmf{dots,label=$_{_{_{n-2}}}$,label.dist=0w,label.side=left}{c1,c3}
    \fmf{dots,label=$_{_{_{n-2}}}$,label.dist=0w,label.side=right}{c4,c5}

\end{fmfgraph*}
\end{center}
\caption{Quantum correction reducible to two point functions}\label{fig:diagrammi funzioni a due punti}
\end{figure}

Technically, to define the two classes of diagrams written in figure \ref{fig:diagrammi funzioni a tre punti},
we must add the diagram in fig. \ref{fig:diagrammi funzioni a due punti},
that for the same reasons as above can be related to the two point function
\begin{equation}
\langle \mathcal O_{L'}^{3} (x_1)\, {\mathcal O^\dagger}^{3}_{R'}(x_3) \rangle \;,
\end{equation}
that has no quantum corrections.

\begin{figure}[b]
\begin{center}
\begin{fmfgraph*}(40,40)
 \fmffreeze
  \fmfpen{.4}

    \fmfforce{0.1w,0.9h}{v1}
    \fmfforce{0.1w,0.9h}{v1a}
    \fmfforce{0.1w,0.1h}{v2}
    \fmfforce{0.1w,0.1h}{v2a}
    \fmfforce{0.9w,0.1h}{v3}
    \fmfforce{0.9w,0.1h}{v3a}
    \fmfforce{0.9w,0.9h}{v4}
    \fmfforce{0.9w,0.9h}{v4a}

    \fmfforce{0.5w,0.85h}{c1}
    \fmfforce{0.5w,0.5h}{c2}
    \fmfforce{0.5w,0.95h}{c3}

    \fmfv{d.sh=circle,d.fill=empty,d.si=2.3thick,l=$x_{1}$,l.a=135}{v1}
    \fmfv{d.sh=cross,d.si=2.3thick}{v1a}
    \fmfv{d.sh=circle,d.fill=empty,d.si=2.3thick,l=$x_{2}$,l.a=-135}{v2}
    \fmfv{d.sh=cross,d.si=2.3thick}{v2a}
    \fmfv{d.sh=circle,d.fill=empty,d.si=2.3thick,l=$x_{4}$,l.a=-45}{v3}
    \fmfv{d.sh=cross,d.si=2.3thick}{v3a}
    \fmfv{d.sh=circle,d.fill=empty,d.si=2.3thick,l=$x_{3}$,l.a=45}{v4}
    \fmfv{d.sh=cross,d.si=2.3thick}{v4a}

    \fmf{fermion,right=.2}{v1,v4}
    \fmf{fermion,left=.2}{v1,v4}
    \fmf{fermion,left=.20}{v1,c2}
    \fmf{fermion,right=.20}{v1,c2}
    \fmf{fermion,left=.20}{c2,v2}
    \fmf{fermion,right=.20}{c2,v2}
    \fmf{fermion,left=.2}{c2,v4}
    \fmf{fermion,right=.2}{c2,v4}
    \fmf{fermion,left=.2}{v3,c2}
    \fmf{fermion,right=.2}{v3,c2}
    \fmfblob{.3w}{c2}
    \fmf{dots,label=$_{_{_{n-2}}}$,label.dist=.001w,label.side=right}{c1,c3}

\end{fmfgraph*}
\qquad\qquad
\begin{fmfgraph*}(40,40)
 \fmffreeze
  \fmfpen{.4}

    \fmfforce{0.1w,0.9h}{v1}
    \fmfforce{0.1w,0.9h}{v1a}
    \fmfforce{0.1w,0.1h}{v2}
    \fmfforce{0.1w,0.1h}{v2a}
    \fmfforce{0.9w,0.1h}{v3}
    \fmfforce{0.9w,0.1h}{v3a}
    \fmfforce{0.9w,0.9h}{v4}
    \fmfforce{0.9w,0.9h}{v4a}

    \fmfforce{0.28w,0.68h}{c1}
    \fmfforce{0.5w,0.5h}{c2}
    \fmfforce{0.32w,0.72h}{c3}
    \fmfforce{0.68w,0.28h}{c4}
    \fmfforce{0.72w,0.32h}{c5}

    \fmfv{d.sh=circle,d.fill=empty,d.si=2.3thick,l=$x_{1}$,l.a=135}{v1}
    \fmfv{d.sh=cross,d.si=2.3thick}{v1a}
    \fmfv{d.sh=circle,d.fill=empty,d.si=2.3thick,l=$x_{2}$,l.a=-135}{v2}
    \fmfv{d.sh=cross,d.si=2.3thick}{v2a}
    \fmfv{d.sh=circle,d.fill=empty,d.si=2.3thick,l=$x_{3}$,l.a=-45}{v3}
    \fmfv{d.sh=cross,d.si=2.3thick}{v3a}
    \fmfv{d.sh=circle,d.fill=empty,d.si=2.3thick,l=$x_{4}$,l.a=45}{v4}
    \fmfv{d.sh=cross,d.si=2.3thick}{v4a}

    \fmf{fermion,right=.3}{v1,c2}
    \fmf{fermion,left=.3}{v1,c2}
    \fmf{fermion,right=.3}{c2,v3}
    \fmf{fermion,left=.3}{c2,v3}
    \fmf{fermion,left=.20}{v1,c2}
    \fmf{fermion,right=.20}{v1,c2}
    \fmf{fermion,left=.20}{c2,v2}
    \fmf{fermion,right=.20}{c2,v2}
    \fmf{fermion,left=.2}{v4,c2}
    \fmf{fermion,right=.2}{v4,c2}
    \fmf{fermion,left=.2}{c2,v3}
    \fmf{fermion,right=.2}{c2,v3}
    \fmfblob{.3w}{c2}
    \fmf{dots,label=$_{_{_{n}}}$,label.dist=.001w,label.side=left}{c1,c3}
    \fmf{dots,label=$_{_{_{n}}}$,label.dist=.001w,label.side=right}{c4,c5}

\end{fmfgraph*}
\end{center}
\caption{Interesting quantum corrections (note the $x_3 \leftrightarrow x_4$ exchange in the first class)}\label{fig:interesting quantum corrections}
\end{figure}
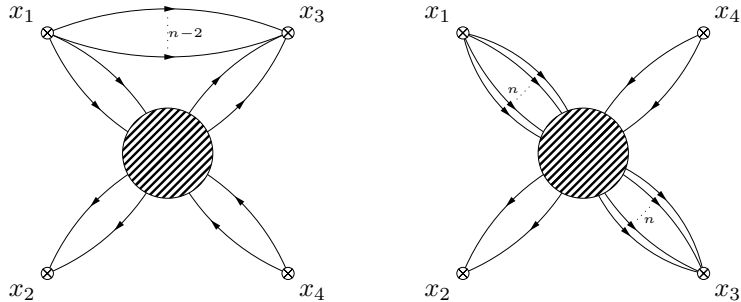

Hence all the non vanishing quantum contributions to the four point function in \eqref{eqn:definizione di Gn} arise from the diagrams
in figure \ref{fig:interesting quantum corrections}.
Actually the second diagram class, depicted as it is, contains all the others, but here we will interpret it as "all the diagrams that remain".
At a generic perturbative order, this class can be very large, but we will see that one can handle it rather easily up to order $g^4$.

We will calculate first $G_n^{(2)}$,
so that we develop the machinery, then the calculation of $G_n^{(1)}$ (and obviously $G_n^{(0)}$) is simpler.

\subsection{Calculation up to two loop}
Let's analyze the first class of diagrams of figure \ref{fig:interesting quantum corrections}.
The diagram is
equivalent to the computation of the correlation function
\begin{equation}
\frac{(n-2)!}{(4 \pi^2)^{n-2}} \frac{1}{(x_{13}^2)^{n-2}}\; Y_{ab|cd}\; \langle
\phi_1^a \phi_1^b(x_1) \, \mathcal C_{11}^\dagger(x_2) \, {\phi_2^\dagger}^c {\phi_2^\dagger}^d(x_3) \, \mathcal C_{22}(x_4) \rangle\;,
\end{equation}
where we encoded in the tensor
\be\label{eqn:definizione di Y}
Y_{ab|cd} \equiv X^L_{a b c_1 \cdots\, c_{n-2}} \,X^R_{c d c_1 \cdots\, c_{n-2}}\;
\ee
all the information about the specific operators we use in the calculation.
This tensor is obviously symmetric in
$a\leftrightarrow b$ and $c\leftrightarrow d$.
The factor $\frac{(n-2)!}{(4 \pi^2)^{n-2}} \frac{1}{(x_{13}^2)^{n-2}}$ is due to the spectator chiral lines connecting the points $x_1$ and $x_3$.
So, for this class of diagrams, the entire
information about the space-time dependence of the perturbative corrections comes from
\begin{equation}\label{eqn:definizione di H}
H^{ab|cd}(x_1,x_2,x_3,x_4) \equiv \langle \phi_1^a \phi_1^b(x_1) \, \mathcal C_{11}^\dagger(x_2) \, {\phi_2^\dagger}^c {\phi_2^\dagger}^d(x_3) \, \mathcal
C_{22}(x_4) \rangle\;.
\end{equation}

Let us analyze now the second class. Here there are all the diagrams that do not belong in one of the other classes.
The interaction bubble must connect \emph{all} the bundles depicted in the tree level diagram (figure \ref{fig:treeleveldiagram}).
This is not possible at order $g^2$, in which the interactions can connect with each other only two chiral lines.
Therefore, at first order in perturbation theory this class is empty.
At two loop it is equivalent to the correlation function
\begin{equation}
\frac{(n-2)^2\,(n-3)!}{(4 \pi^2)^{n-3}} \frac{1}{(x_{13}^2)^{n-3}}\; Z_{abc|de\!f}\; L^{abc|de\!f}(x_1,x_2,x_3,x_4)\;,
\end{equation}
where
\begin{gather}\label{eqn:definizione di Z}
Z_{abc|de\!f}\equiv X^L_{a b c \,c_1 \cdots\, c_{n-3}} \,X^R_{d e \!f c_1 \cdots\,c_{n-3}}\;,\\
\begin{split}
L^{abc|de\!f}(x_1,x_2,x_3,x_4)&\equiv \\\langle \phi_1^a \phi_1^b \phi_3^c(x_1) &\, \mathcal C_{11}^\dagger(x_2) \, {\phi_2^\dagger}^d {\phi_2^\dagger}^e
{\phi_3^\dagger}^f(x_3) \, \mathcal C_{22}(x_4) \rangle
- \frac{\delta^{cf}}{(4 \pi^2) x_{13}^2} H^{ab|de}(x_1,x_2,x_3,x_4)\label{eqn:definizione di L}
\;.
\end{split}
\end{gather}
The term subtracted in eq. \eqref{eqn:definizione di L} belongs to the first diagram class in figure \ref{fig:interesting quantum corrections}.
Note that the relation $\delta^{cf}Z_{abc|de\!f}=Y_{ab|de}$ holds.

We have reduced the calculation of a generic four point function $G_n$ at order $g^4$ to the calculation of two
particular functions\footnote{These are \emph{not} correlators of well defined conformal operators,
thus we do not have to expect them to be conformal invariant.},
$H^{ab|cd}$ and $L^{abc|de\!f}$. In these functions all the information about the space-time dependence is encoded.
The other ($g$ independent) information is in the two color tensors $Y_{ab|cd}$ and $Z_{abc|de\!f}$.
To calculate $G_n^{(2)}$ we have to analyze the different color structures of the two functions $H^{ab|cd}$ and
$L^{abc|de\!f}$. Contracting these structures with the color tensors $Y_{ab|cd}$ and $Z_{abc|de\!f}$ will provide the full result.

\subsubsection{Color structures at order $g^4$}
The computation of the two functions $H^{ab|cd}$ and $L^{abc|de\!f}$ described in (\ref{eqn:definizione di H}, \ref{eqn:definizione di L}) requires the calculation of many diagrams,
which are drawn in the appendix.
The functions have free color indices, so each of the Feynman diagrams is multiplied by a precise color structure.
This structure has to be contracted with the color tensors $Y_{ab|cd}$ or $Z_{abc|de\!f}$, depending on the class to which the diagram belongs.

In table \ref{tab:color structures and diagrams} we list the color structures we find, and the diagrams that provide it.
\begin{table}[htbp]
\begin{center}
\begin{tabular}{|c|c|}
\hline
Structure & Diagrams\\
\hline
\hline
$H_1^{ab|cd} = f_{i a c} f_{i b d}$ &
$B_{{2}}\,,B_{{3}}\,,B_{{4}}\,,B_{{5}}\,,B_{{7}}\,,B_{{9}}\,,B_{{10}}\,,B_{{11}}\,,B _{{13}}\,,B_{{14}}\,,B_{{19}}\,,B_{{20}}$\\
\hline
$H_2^{ab|cd} = f_{i k a} f_{j k b} f_{j \ell c} f_{i \ell d}$&
$A_1\,, A_4\,,B_{{1}}\,,B_{{6}}\,,B_{{8}}\,,B_{{12}}\,,
C_1\,,C_2\, \cdots C_9
$\\
\hline
$H_3^{ab|cd} = f_{i \ell a} f_{j k b} f_{j \ell c} f_{i k d}$&
$B_{{15}}\,,B_{{16}}\,,B_{{17}}\,,B_{{18}}\,,B_{{21}}$\\
\hline
$H_4^{ab|cd} = f_{\ell a c} f_{j k b} f_{i k d} f_{i j \ell}$ &
$A_2\,, A_3$\\
\hline
$L_1^{abc|de\!f} = f_{k a e} f_{j b \!f} f_{i j c} f_{i k d}$&
$A_5$\\
\hline
$L_2^{abc|de\!f} = f_{j a e} f_{k c \!f} f_{i k b} f_{i j d}$&
$B_{22}\,, B_{23}\,, B_{24}\,, B_{25}\,, B_{26}\,, B_{27}$\\
\hline
\end{tabular}
\end{center}
\caption{Color structures and two loop diagrams}\label{tab:color structures and diagrams}
\end{table}
Symmetrization of the color indices is not taken into account.
This is guaranteed by the contraction with the tensors
$Y_{ab|cd}$ and $Z_{abc|de\!f}$.
Taking this symmetrization into account we can relate these structures.
In fact:
\begin{align}
Y_{ab|cd}\, H_3^{ab|cd} &=Y_{ab|cd}\left( H_2^{ab|cd} - \frac{N}{2} H_1^{ab|cd}\right)\;, & Y_{ab|cd}\, H_4^{ab|cd} &= - \frac{N}{2} Y_{ab|cd}\, H_1^{ab|cd}\;.
\end{align}

\subsubsection{Two loop result}
Given these formulae, one can write down the complete result at order $g^4$ for the two functions.
For the function $H^{ab|cd}$ we have
\begin{multline}
\begin{split}
\left.H^{ab|cd}\right|_{g^4}= g^4\,N \Bigl[ 8(A_2 + A_3)
&- 16(2\, B_2 +B_3 +B_4 -  B_5 +2\, B_7 + B_9 +B_{10} -B_{11}\\& +B_{13} +B_{14} -B_{15} -B_{16} -B_{17}+B_{18}-B_{19}-B_{20}+B_{21})\Bigr] H_1^{ab|cd}
\end{split}
\\
\begin{split}
+ g^4 \Bigl[ 8 (2\, A_1+ A_4)& -32 (B_1 -B_6 +B_8 -B_{12} +B_{15} +B_{16} + B_{17} + B_{18} -B_{21}) \\
&+32 (2\,C_1 +2\, C_2-2\,C_3-2\,C_4 +C_5 -C_6+2\,C_7+2\,C_8-C_9)
\Bigr] H_2^{ab|cd}\;.
\end{split}
\end{multline}
The function $L^{abc|de\!f}$ is
\be
\left. L^{abc|de\!f}\right|_{g^4}=g^4 \Bigl[16 \, A_5
\Bigr] L_1^{abc|de\!f}\;.
\ee
The structure that multiplies $L_2$ is absent, since the linear combination of diagrams $B_{22}$ to $B_{27}$ which multiplies it sums to zero
(see table \ref{tab:color structures and diagrams} and appendix).

Contracting these structures with the tensors $Y_{ab|cd}$ and $Z_{abc|de\!f}$ (coming from the contraction of the color tensors $X^L$ and $X^R$,
see eqs. \eqref{eqn:definizione di Y}, \eqref{eqn:definizione di Z}) we get the full result, which will depend upon the contractions
\be\label{eqn:contrazioni indipendenti}
Y \cdot H_1\equiv Y_{ab|cd} H_1^{ab|cd}\,, \qquad Y \cdot H_2\equiv Y_{ab|cd} H_2^{ab|cd}\,, \qquad Z \cdot L_1\equiv Z_{abc|de\!f} L_1^{abc|de\!f}\;.
\ee
Putting everything together we obtain (see appendix)
\begin{multline} \label{eqn:risultato di G2}
G_n^{(2)}(x_1,x_2,x_3,x_4)= \frac{N}{(4 \pi^2)^{n+4}} \frac{(n-2)!}{(x_{13}^2)^{n-2}x_{12}^4 x_{34}^4}\Biggl\{
\frac{r}{4} B\left(r,s\right)^2 \left[(Y \cdot H_1) \left( r + s - 1 \right) +
2\, \frac{(Y \cdot H_2)}{N} \right] \Biggr.\\ \Biggl. + (Y \cdot H_1) \left[
\Phi^{(2)}\left(\frac{1}{r},\frac{s}{r}\right) +  \frac{r}{s} \Phi^{(2)}\left(\frac{1}{s},\frac{r}{s}\right) \right] +
r\, \Phi^{(2)}\bigl(r,s\bigr)\left[(Y \cdot H_1) + (n-2) \frac{(Z \cdot L_1)}{N} \right]\Biggr\}\;,
\end{multline}
where $r$ and $s$ are the conformally invariant cross ratios
\be\label{eqn:definizione di r e s}
r=\frac{x_{12}^2 x_{34}^2}{x_{13}^2 x_{24}^2}\;,
\quad
s=\frac{x_{14}^2 x_{23}^2}{x_{13}^2 x_{24}^2}\;.
\ee

\subsubsection{Lower order computations}
We performed the computation at order $g^4$. One can make exactly the same considerations to calculate these
functions at order $g^2$. As stressed before, we can do this using simply the first class of
diagrams in figure \ref{fig:interesting quantum corrections}, since the second does not contribute.
In fact only one Feynman diagram will appear in the computation, multiplied by
the structure $H_1$. So
\be \label{eqn:risultato di G1}
G_n^{(1)}(x_1,x_2,x_3,x_4)=-\frac{2}{(4 \pi^2)^{n+3}} \frac{(n-2)!}{(x_{13}^2)^{n-2}x_{12}^4 x_{34}^4} r B\left(r,s\right) (Y \cdot H_1)\;.
\ee
Hence to calculate completely the function we need only the color tensor $Y_{ab|cd}$.

Similarly, at tree level we find\footnote{We
can use instead of $Y_{ab|cd}$ the tensor $Z_{abc|de\!f}$, because we know that $\delta^{cf} Z_{abc|de\!f}=Y_{ab|de}$.
Our notation is more general, because when $n=2$ the tensor $Z$ does not exist.}

\be \label{eqn:risultato di G0}
G_n^{(0)}(x_1,x_2,x_3,x_4)=\frac{(n-2)!}{(4 \pi^2)^{n+2}} \frac{Y_{ab|cd} \delta^{ab} \delta^{cd}}{(x_{13}^2)^{n-2}x_{12}^4 x_{34}^4}\;.
\ee

\subsection{Application to more general operators}
All these considerations were done for $\frac{1}{2}$ BPS operators.
To neglect the contributions coming from the diagrams written in figure \ref{fig:diagrammi funzioni a tre punti},
we used the strong non renormalization theorem for the extremal three point functions of $\frac{1}{2}$ BPS operators \cite{nonrenorm},
valid at all orders in perturbation theory.
Hence, for $\frac{1}{2}$ BPS operators we can apply the same reasoning at all orders in $g$.

At order $g^4$ our results actually can be extended to a more general class of operators.
Our considerations are based on the total symmetry of the color tensors $X^{L,R}$ defining the operators in external lines.
A careful inspection shows that all the results would be the same provided that $X^{L,R}=X^{L,R}_{(a_1 a_2) (a_3\cdots a_n)}$,
i.e. with the symmetry only in the first two and in the last $n-2$ indices.
Note that with our flavour choice this is always guaranteed.
Thus one has only to check whether the operator has such a flavour representative.

In order to apply our method, the contributions coming from the diagrams in figure
\ref{fig:diagrammi funzioni a tre punti}, must
have neither quantum corrections at order $g^2$, nor logarithmic divergencies at order $g^4$.
So, if the operators $\mathcal{O}_{L,R}$ have vanishing one and two loop anomalous dimensions,
we only need to check the vanishing of the order $g^2$ contribution to the three point functions
\be \label{eqn:correlatori che funzionano lo stesso}
\langle{\mathcal O}_{L'}  \mathcal{Q}_\mathbf{20'} \mathcal{O}_{R} \rangle\;, \qquad \langle{\mathcal O}_{L}  \mathcal{Q}_\mathbf{20'} \mathcal{O}_{R'} \rangle\;.
\ee
Here $\mathcal O_{L',R'}$, coming from the contraction of two scalar fields that
belongs to the original $ \mathcal{O}_{L,R}$ are always $\frac{1}{2}$ BPS operators due
to our flavour choice (see eqs. \eqref{eqn:definizione Ol'}, \eqref{eqn:definizione di Or'}).
This extends our technique also to other protected operators.
It has been demonstrated that the three point functions of two $\frac{1}{2}$ BPS
with one general BPS operator do not receive order $g^2$ corrections \cite{D'Hoker:2001bq}.
Hence our result can be extended to all the BPS operator that can be written as indicated in \eqref{eqn:representatives}.
We will give in section \ref{sec:examples} an explicit example involving a $\frac{1}{4}$ BPS operator.

\section{OPE analysis of the four point function} \label{sec:general OPE expansions}
One can use the four point functions calculated in such a way to extract information about the theory.
As an illustration, here we perform the OPE of the function in the OPE s-channel,
in order to study the different contributions coming from the conformal families of operators that appear in the expansion.
For this purpose we can consider
\be\label{eqn:definizione di limite}
G_n(x_2,x_3,x_4)=\lim_{x_1 \rightarrow \infty} x_1^{2n}\, G_n(x_1,x_2,x_3,x_4)\;,
\ee
and expand this function in the limit $x_4 \rightarrow x_3$.
In this limit the final answer will depend on the three variables $x_{23}^2$, $x_{24}^2$, $x_{34}^2$,
simplifying the space time dependence.

This expansion will have logaritmic singularities, written in terms of the expansion parameter $x_{34}$.
In the OPE interpretation,
the coefficients of these singularities can be interpreted as coming from the operators passing in the intermediate OPE channel,
in particular their trilinear couplings (up to order $g^2$) and their one and two loop anomalous dimensions.
If we have enough four point functions, these quantities can be obtained solving a system coming from the coefficients of the expansions.

The expansion of $G_n^{(0)}$ does not have logarithmic singularities. It is simply
\be\label{eqn:espansione a tree level}
G_n^{(0)}(x_2,x_3,x_4)=\frac{(n-2)!}{(4 \pi^2)^{n+2}} \frac{Y_{ab|cd} \delta^{ab} \delta^{cd}}{x_{34}^4}\;.
\ee

We shall write down the order $g^2$ and $g^4$ contributions, namely the functions $G_n^{(1)}$ and $G_n^{(2)}$,
that involve the functions $B$ and $\Phi^{(2)}$, given in the appendix.
We start with $G_n^{(1)}$
\begin{multline}\label{eqn:espansione a 1 loop}
G_n^{(1)}(x_2,x_3,x_4)= \frac{(n-2)!}{(4 \pi^2)^{n+3}}\frac{Y \cdot H_1}{x_{34}^2 x_{23}^2}\Biggl\{
\Bigl[-4 - 2\,\Lambda\Bigr] +\Bigl[2+2\,\Lambda \Bigr]\frac{x_{23} \cdot x_{34}}{x_{23}^2}\\+\Bigl[-\frac{16}{9} -
\frac{8}{3}\,\Lambda \Bigr]\frac{(x_{23} \cdot x_{34})^2}{x_{23}^4}
+ \Bigl[\frac{4}{9} + \frac{2}{3} \Lambda \Bigr] \frac{x_{34}^2}{x_{23}^2}+\cdots
\Biggr\}\;,
\end{multline}
where the dots stand for higher order contributions in the expansion parameter $x_{34}$,
and $\Lambda=\frac{\log(x_{23}^2)}{\log(x_{34}^2)}$.
There are two different expansions nested within each other: the $x_{34}$ expansion,
from which we recognize the \emph{bare} dimension and the Lorentz structure of the operators that appear (see \cite{ratio}),
and the $\Lambda$ expansion, that allows us to extract different quantities, namely the trilinear couplings and/or the anomalous dimensions.

For the two loop contribution we get
\begin{multline}\label{eqn:espansione a 2 loop}
G_n^{(2)}(x_2,x_3,x_4)=\frac{(n-2)!}{(4 \pi^2)^{n+4}}\frac{1}{x_{34}^2 x_{23}^2}\Biggl\{\\
\begin{split}
\Biggl[\text{finite term} +&
\Bigl(6\,N (Y \cdot H_1) + 2 (Y \cdot H_2) + 3 (n-2) (Z \cdot L_1) \Bigr) \Lambda +\\&
\Bigl(N (Y \cdot H_1) + \frac{1}{2} (Y \cdot H_2) +\frac{1}{2} (n-2) (Z \cdot L_1) \Bigr) \Lambda^2
\Biggr]
\end{split}\\
\begin{split}
+\Biggl[\text{finite term} +&
\Bigl(-2\,N (Y \cdot H_1) -3 (Y \cdot H_2) -\frac{13}{4} (n-2) (Z \cdot L_1) \Bigr) \Lambda +\\&
\Bigl(-\frac{1}{2} N (Y \cdot H_1) -(Y \cdot H_2) -\frac{3}{4} (n-2) (Z \cdot L_1) \Bigr) \Lambda^2
\Biggr]\frac{x_{23} \cdot x_{34}}{x_{23}^2}\end{split}\\
\begin{split}
+\Biggl[\text{finite term} +&
\frac{1}{9}\Bigl(17\,N (Y \cdot H_1) +41 (Y \cdot H_2) +40 (n-2) (Z \cdot L_1) \Bigr) \Lambda +\\&
\Bigl(\frac{4}{9} N (Y \cdot H_1) +\frac{11}{6}(Y \cdot H_2) +\frac{11}{9} (n-2) (Z \cdot L_1) \Bigr) \Lambda^2
\Biggr]\frac{(x_{23} \cdot x_{34})^2}{x_{23}^4}\end{split}\\
\begin{split}
+\frac{1}{36}\Biggl[\text{finite term} +&
\Bigl(19\,N (Y \cdot H_1) -32 (Y \cdot H_2) -49 (n-2) (Z \cdot L_1) \Bigr) \Lambda +\\&
\Bigl(5 N (Y \cdot H_1) -12(Y \cdot H_2) -11 (n-2) (Z \cdot L_1) \Bigr) \Lambda^2
\Biggr]\frac{x_{34}^2}{x_{23}^2}+\cdots\Biggr\}\end{split}
\end{multline}
We do not write explicitly the finite terms because their OPE interpretation contains also the $g^4$ part of the trilinear couplings,
that will introduce new unknowns in our system.
In general, to extract this quantity we need also other equations coming from three loop calculations \cite{Bianchi:2003eg}.

From the leading terms in the expansion parameter (\ref{eqn:espansione a 1 loop}, \ref{eqn:espansione a 2 loop}),
one can see that in this channel the lowest dimensional operators that contribute in a non trivial way\footnote{Namely with
some quantum corrections, either in their scaling dimension or trilinear couplings.}
have bare dimension $\Delta^{(0)}=n$ and are Lorentz scalars.
The lowest dimensional rank 2 tensors that appear have bare dimension $\Delta^{(0)}=n+2$, due to the $\frac{(x_{23} \cdot x_{34})^2}{x_{34}^2  x_{23}^6 }$ terms.
We will see at the end of this section how it is possible to extract information also from these considerations.

\subsection{OPE interpretation: the lowest dimensional intermediate operators}
As an example, we focus our attention to the operators of bare dimension $n$ that appear in the OPE s-channel of this correlators.
Their contribution is associated to the leading terms in the expansion parameter $x_{34}$ in
\eqref{eqn:espansione a 1 loop}, \eqref{eqn:espansione a 2 loop},
from which we can infer that they are all Lorentz scalars.
Hence, in the most general case, a set of Lorentz scalars operators $\mathcal O_i$ of $\Delta^{(0)}=n$ can appear.
The OPE structure constants of the operators in the external points ${C_{L q}}^i$ and ${C_{R q}}^i$ are defined as
\be\label{eqn:normalizzazioni e ope structure constants}
{C_{L,R q}}^i(g) \equiv \frac{C_{L,R q i}(g)}{M_i(g)}\;,
\ee
where $C_{L,R q i}$ are the trilinear couplings with the operators in the external points and $M_i$ are
the two point function normalization of the intermediate operators:
\begin{align}
\langle\mathcal{O}_{L,R}(x)\; \mathcal Q_\mathbf{20'}(y)\; \mathcal O_i(0)  \rangle &=
\frac{C_{L,R q i}(g)}{x^{2n-2}\,(x-y)^2\, y^2}
\left(\frac{(x-y)^2}{x^2\, y^2}\right)^{\frac{1}{2}\left(\Delta_i(g^2) - n\right)}\;,\\
\langle \mathcal O_i(x)\,\mathcal O_i(0)\rangle&=
\frac{M_i(g)}{(x^2)^{\Delta(g^2)}}\;.\label{eqn:funzioni a due punti degli operatori intermedi}
\end{align}
The coefficients that appear in the OPE analysis do not depend the normalization of the intermediate operators.
We can choose $M_i(g)= 1$, so that all the $g$-dependence
of the two point function \eqref{eqn:funzioni a due punti degli operatori intermedi} comes only from the anomalous dimension.
For this particular normalization the OPE structure constants
coincide with the trilinear couplings. If we want to recover the latter for different normalizations we need to take into account
equation \eqref{eqn:normalizzazioni e ope structure constants}.

From the expansions (\ref{eqn:espansione a tree level}, \ref{eqn:espansione a 1 loop}, \ref{eqn:espansione a 2 loop})
and the OPE interpretation \cite{ratio}, putting
\begin{align}
\Delta_i(g^2)&=n +g^2 \gamma^{(1)}_i+ g^4 \gamma^{(2)}_i+ \cdots\;,\\
C_{L,R q i}(g)&=C^{(0)}_{L,R q i}+ g^2 C^{(1)}_{L,R q i}+\cdots\;,
\end{align}
 we have
\begin{gather}
\sum_i C^{(0)}_{L\mathcal Q i} C^{(0)}_{R\mathcal Q i} = 0\,, \label{eqn:sommatreelevel}
\\
-\frac{1}{2} \sum_i C^{(0)}_{L\mathcal Q i} C^{(0)}_{R\mathcal Q i} \gamma^{(1)}_i =
- \frac{2 (n-2)!}{(4 \pi^2)^{n+3}} (Y \cdot H_1)\,,
\\
 \sum_i \left(C^{(0)}_{L\mathcal Q i} C^{(1)}_{R\mathcal Q i}+ C^{(1)}_{L\mathcal Q i} C^{(0)}_{R\mathcal Q i}\right) =
- \frac{4 (n-2)!}{(4 \pi^2)^{n+3}} (Y \cdot H_1)\,,
\\
\frac{1}{8} \sum_i C^{(0)}_{L\mathcal Q i} C^{(0)}_{R\mathcal Q i} {\gamma^{(1)}_i}^2 =
 \frac{(n-2)!}{(4 \pi^2)^{n+4}}\Bigl(N (Y \cdot H_1) + \frac{1}{2} (Y \cdot H_2) +\frac{1}{2} (n-2) (Z \cdot L_1) \Bigr)\,,
\\
\begin{split}
-\frac{1}{2} \sum_i \Bigl[
\bigl(C^{(0)}_{L\mathcal Q i} C^{(1)}_{R\mathcal Q i}+ C^{(1)}_{L\mathcal Q i} C^{(0)}_{R\mathcal Q i}\bigr)\gamma^{(1)}_i
&+  C^{(0)}_{L\mathcal Q i} C^{(0)}_{R\mathcal Q i} \gamma^{(2)}_i \Bigr] =\\
=& \frac{(n-2)!}{(4 \pi^2)^{n+4}}\Bigl(6\,N (Y \cdot H_1) + 2 (Y \cdot H_2) + 3 (n-2) (Z \cdot L_1) \Bigr)\,.\label{eqn:somma2loop}
\end{split}
\end{gather}
We spend a few words about equation \eqref{eqn:sommatreelevel}. The rhs
is equal to zero because in eq. \eqref{eqn:espansione a tree level} we do not have a $\frac{1}{x_{34}^2 x_{23}^2}$ term.
This is the unique equation in which will also contribute protected operators whose trilinear couplings do not renormalize.
These operators do not appear at all in the other equations.
Their contributions can be easily derived from a simple tree level calculation.

\subsection{Information from the OPE}
The machinery we developed allows us to write a system in which the unknowns are the anomalous dimensions of the operators that appear in the OPE channels,
and the trilinear couplings of these operators
(for a given choice of the normalization of their two point function)
with the operators in the external points.
If we have enough four point functions, this system can be solved.
The number of necessary four point functions depends on how many operators with the same bare dimension appear in that channel.
As we know, this is determined by the fusion rules, which must obey also the constraints imposed by group theory.
The representation to which belong the operators in the intermediate channel must be in the product of the representation of the external operators
that select the OPE channel.

One can extract information also in the opposite direction: inspecting the OPE expansion of $G_n$ there are some operators
whose presence in the OPE channel is not forbidden by group theory, but they do not appear beyond tree level, or do not appear at all.
We saw that in the s-channel there are no scalar fields (in the interacting regime) up to bare dimension $n-2$,
and no rank 2 tensor fields up to bare dimension $n$.
When there are
only $\frac{1}{2}$ BPS operators in the external points,
the former is a consequence of the general non renormalization theorem for extremal three point functions.
In fact one can see that in the product of $[0,2,0]$ and $[0,n,0]$ at bare dimension $\Delta^{(0)}=n-2$ only the
representation $[0,n-2,0]$ is allowed, hence in the internal channel only $\frac{1}{2}$ BPS operators will appear.
It is worthwhile to ask what is the particular set of operators that appears in our OPE expansion, since
we can infer useful information also from the fields that do not appear.

Before generalizing the considerations of this kind,
we have to remind that we made a precise choice of the flavour representatives of the operators in the external points.
This can project out some operators, that can appear for different external representatives.
For an illustration of this see for example \cite{Arutyunov:2001mh}.
General conclusions can be drawn with the help of this method only after having checked that the absence of some field is not due to the
particular flavour choice.

\section{Examples}\label{sec:examples}
In this section we will perform the described calculation for the four point functions of the type $G_3$ and $G_4$ (see eq. \eqref{eqn:definizione di Gn}),
and see how these results can be used to extract trilinear couplings and anomalous dimensions of the operators that appear in the OPE channels.
In particular we will study also a case in which the operators in the external points are not all $\frac{1}{2}$ BPS.
In the case of only $\frac{1}{2}$ BPS operators we calculate $G_n$ also for $n=5,6$. These results are given in the appendix.

\subsection{A case without mixing: the operator  $\mathcal{S}_{\mathbf{6}}$}
A very lucky case is the one in which there is only one operator with a given bare dimension.
In this situation we need only one four point function to obtain the desired coefficients.

Here we will analyze the OPE expansion of the function $G_3$, from which we can extract information
about the operator \be
\mathcal{S}_{\mathbf{6}}^J \equiv \sum_{I=1}^3 \text{tr}
\left(\phi_{I}\phi^\dagger_{I} \phi_{J}\right)
\;.
\ee
This is a scalar operator of bare
dimension $\Delta^{(0)}=3$, belonging to the representation
$\mathbf{6}$ of the flavour group.

\subsubsection{The four  point function}
Let us analyze the correlator \be G_3 (x_1,x_2,x_3,x_4) \equiv \langle
\text{tr} ( \phi_1 \phi_1 \phi_3)(x_1)\; \mathcal{C}^\dagger_{11}(x_2) \; \text{tr} ( \phi_2^\dagger \phi_2^\dagger
\phi_3^\dagger)(x_3) \;\mathcal{C}_{22}(x_4)\rangle \;.\ee This is a correlation function of two operators
$\mathcal{Q}_{\mathbf{20'}}$ with two $\frac{1}{2}$ BPS operators, having $\Delta=3$ in the representation $[0,3,0]=\mathbf{50}$
of the flavour group. We call these operators $\mathcal{Q}_{\mathbf{50}}$.
In the language of eq. \eqref{eqn:representatives} we have
\begin{align}
\mathcal Q_\mathbf{50}^{11;3} &\equiv \frac{1}{4} d_{a b c} \phi_1^a \phi_1^b \phi_3^c\;, &
{\mathcal Q_\mathbf{50}^\dagger}^{22;3} &\equiv \frac{1}{4} d_{a b c} {\phi_2^\dagger}^a {\phi_2^\dagger}^b {\phi_3^\dagger}^c\;,
\end{align}
where $d_{abc}$ is the totally symmetric tensor coming from the anticommutator of two generators of the color group $SU(N)$, according to
\be
\left\{ T_a , T_b \right\} = \delta_{ab}\frac{1}{N} + d_{abc} T_c\;.
\ee

In the OPE between the operators $\mathcal{Q}_{\mathbf{50}}(x_3)$ and $\mathcal{Q}_{\mathbf{20'}}(x_4)$ we can find
also the operator $\mathcal{S}_\mathbf 6$. In fact from group theory we have that
$$
\mathbf{20'}\times \mathbf{50} = \mathbf{6} + \mathbf{50} + \mathbf{64} + \mathbf{196} + \mathbf{300} + \mathbf{384}\;.
$$
We know that the first non trivial contributions to the OPE expansions of the functions $G_n$ come from operator of bare dimension $\Delta^{(0)}=n$,
so the contribution coming from $\mathcal{S}_\mathbf 6$ is the leading one in the
expansion in $x_{34}$. There are no other unprotected operators of bare dimension 3 belonging to the allowed representations.
The other operator with such characteristics is $\mathcal{Q}_{\mathbf{50}}$, that does not contribute beyond tree level.
Then $G_3$ meets all the necessary requirements in order to
calculate the anomalous dimension and the trilinear couplings of $\mathcal{S}_\mathbf 6$.

As we saw before, to calculate the function $G_3$ we need to construct the color tensors $Y_{ab|cd}$ and $Z_{abc|de\!f}$
and contract them with the three independent color structures given above, see eq. \eqref{eqn:contrazioni indipendenti}.
For the given four point function we find
\begin{align}
Y_{ab|cd}&=\frac{1}{16} d_{abe}d_{cde} & Z_{abc|de\!f}&=\frac{1}{16} d_{abc}d_{de\!f} \\
Y_{ab|cd}\delta^{ab}\delta^{cd}&=0 & Y\cdot H_1 &= \frac{(N^2-1)(N^2-4)}{32} \\
Y\cdot H_2 &= \frac{N(N^2-1)(N^2-4)}{64} &  Z\cdot L_1 &= -\frac{N(N^2-1)(N^2-4)}{64}\;,
\end{align}
so
\begin{align}
G_3^{(0)}(x_1,x_2,x_3,x_4)=&\;0\;,\\
G_3^{(1)}(x_1,x_2,x_3,x_4)=&-\frac{(N^2-1)(N^2-4)}{16(4 \pi^2)^{6}} \frac{r}{x_{13}^2 x_{12}^4 x_{34}^4} B\left(r,s\right)\;,\\
G_3^{(2)}(x_1,x_2,x_3,x_4)=&\frac{N(N^2-1)(N^2-4)}{32(4 \pi^2)^{7}} \frac{1}{x_{13}^2 x_{12}^4 x_{34}^4}
\Biggl\{\frac{r}{4} B\left(r,s\right)^2
\left[r + s \right] \\
& +
\left[\Phi^{(2)}\left(\frac{1}{r},\frac{s}{r}\right) +  \frac{r}{s} \Phi^{(2)}\left(\frac{1}{s},\frac{r}{s}\right)  +\frac{1}{2}
r\, \Phi^{(2)}\bigl(r,s\bigr) \right] \Biggr\}\;,
\end{align}
where $r$ and $s$ are defined in \eqref{eqn:definizione di r e s}.

\subsubsection{Anomalous dimensions and trilinear couplings}
After performing the expansion of the computed functions and the OPE interpretation
we obtain the one and two loop anomalous dimensions and the trilinear couplings of
$\mathcal S_\mathbf 6$ with $\mathcal Q_\mathbf{50}$ and $\mathcal Q_\mathbf{20'}$.
From the general expansions we can extract
\begin{align}
C^{(0)}_{\mathcal Q_\mathbf{50} \mathcal Q_\mathbf{20'} \mathcal S_\mathbf{6}} &=
\frac{1}{4(4 \pi^2)^4}\frac{(N^2-1)(N^2-4)}{N} &
C^{(1)}_{\mathcal Q_\mathbf{50} \mathcal Q_\mathbf{20'} \mathcal S_\mathbf{6}} &=
\frac{1}{4(4 \pi^2)^5}(N^2-1)(N^2-4) \\
\gamma_{\mathcal S_\mathbf{6}}^{(1)}&= \frac{2 N}{4 \pi^2} &
\gamma_{\mathcal S_\mathbf{6}}^{(2)}&= -\frac{3 N^2}{2(4 \pi^2)^2} &
\end{align}
Here the operator $\mathcal S_\mathbf 6$ is normalized such that:
\be
\langle \mathcal S_\mathbf 6(x_1)\,\mathcal S_\mathbf 6(x_2)\rangle=
\frac{1}{(4 \pi^2)^3} \frac{(N^2-1)(N^2-4)}{N}
\frac{1}{(x_{12}^2)^{\Delta}}\;,
\ee
where $\Delta=3+g^2 \gamma_{\mathcal S_\mathbf{6}}^{(1)}+ g^4 \gamma_{\mathcal S_\mathbf{6}}^{(2)}+\cdots$.
For simplicity here we understood the insertion of the wave function renormalization.
The anomalous dimensions derived here match \cite{Beisert:2003jj}.

\subsection{A case with mixing: correlators of type $G_4$}
Let us consider the functions $G_4$. The representation $[0,4,0]$ of $SU(4)$ has dimension 105.
There are two operators belonging to this representation with dimension 4, that we call $\mathcal D_\mathbf{105}$
and $\mathcal Q_\mathbf{105}$, respectively the double trace and the single trace operator:
\begin{align}
\mathcal D_\mathbf{105}^{1133} &\equiv \left(\delta_{ab} \delta_{cd} + \delta_{ac} \delta_{bd}+ \delta_{ad} \delta_{bc} \right)
\;\phi_1^a \phi_1^b \phi_3^c \phi_3^d\;,\\
\mathcal Q_\mathbf{105}^{1133} &\equiv \left(\text{tr}\left(T_a T_b T_c T_d \right) + \text{symmetrization} \right)
\;\phi_1^a \phi_1^b \phi_3^c \phi_3^d\;.
\end{align}
So there are three four point function in this class, that we call
$G_{\cal DD}$,  $G_{\cal QQ}$ and $G_{\cal DQ}$.

At dimension $\Delta=4$ there is also another BPS operator for which this method can be applied.
This is the $\frac{1}{4}$ BPS operator $\mathcal D_\mathbf{84}$.
In fact, one can easily verify that
\be
\mathcal D_\mathbf{84}^{1133} \equiv \left(\delta_{ab} \delta_{cd} - \delta_{ac} \delta_{bd} -\frac{2}{N} f_{a c e} f_{b d e}\right)
\phi_1^a \phi_1^b \phi_3^c \phi_3^d
\ee
is a good representative of this operator.
So we can use for it our method, and build the correlators $G_{d \cal D}$, $G_{d \cal Q}$ and $G_{d d}$,
where $\mathcal D_\mathbf{84}$ is represented with $d$.
Also the protected operator $\mathcal D_\mathbf{20'}$ satisfies the same properties as $\mathcal D_\mathbf{84}$,
but we cannot use for it our method because the correct definition of this operator is less trivial, and has also $g$ - dependent corrections
\cite{Bianchi:2003eg}.

In this way we can calculate the relevant contributions to the six four point functions under consideration.
So we get the numbers written in table \ref{Color contraction for functions of $G_4$ type}.
\begin{table}[hbtp]
\begin{center}
\begin{tabular}{||c||c|c||}
\hline
\hline
Function & $Y\cdot \delta^2$ & $Y \cdot H_1$  \\
\hline
$G_{\cal DD} $&$(N^4-1)(N^2+1)$   & $N (N^4-1)$                                  \\
$G_{\cal DQ}$ &$\frac{1}{2N} (N^4-1)(2 N^2-3)$   & $\frac{1}{2} (N^2-1)(2 N^2-3)        $ \\
$G_{\cal QQ}$ &$\frac{1}{4N^2} (N^2-1)(2 N^2-3)^2$   & $\frac{1}{8N} (N^2-1)(N^4 -6 N^2 +18)$ \\
$G_{dd}     $ &$(N^2-1)(N^2-4)^2$   & $N (N^2-1) (N^2-4)                   $ \\
$G_{d\cal D}$ &$(N^4-1)(N^2-4)$   & $N (N^2-1) (N^2-4)                   $ \\
$G_{d\cal Q}$ &$\frac{1}{2N} (N^2-1)(N^2-4)(2 N^2-3)$   & $\frac{3}{4} (N^2-1)(N^2-4)          $ \\
\hline
\hline
Function  & $Y\cdot H_2$ & $Z \cdot L_1$\\
\hline
$G_{\cal DD} $  &$ N^2 (N^2-1) (N^2+6) $               & $\frac{5}{2} N^2 (N^2-1)$        \\
$G_{\cal DQ}$  &$ \frac{5}{4} N^3 (N^2-1)          $ &$ \frac{1}{8} N (N^2-1) (N^2+6) $ \\
$G_{\cal QQ}$ &$ \frac{1}{16} N^2 (N^2-1)(N^2+6)$ &$-\frac{1}{32} (N^2-1)(N^4 -18 N^2 +36)$\\
$G_{dd}     $  &$ \frac{1}{2}(N^2-1)(N^2-4)(2 N^2-3)$ &$-3 (N^2-1) (N^2-4) $\\
$G_{d\cal D}$  &$ N^2 (N^2-1) (N^2-4)               $ &$ 0$ \\
$G_{d\cal Q}$  &$ \frac{3}{4} N (N^2-1)(N^2-4)      $ &$ 0$ \\
\hline
\hline
\end{tabular}
\end{center}
\caption{Color contractions for functions of type $G_4$}\label{Color contraction for functions of $G_4$ type}
\end{table}

\subsubsection{{Scalar operators of $\Delta^{(0)}=4$ in the representation $\mathbf{20'}$}}
These four point functions contain enough information to solve the system coming from the OPE expansion,
in which the operators that pass are the three operators $\mathcal O_i$ written in \cite{Bianchi:2002rw}.
In fact, group theory gives:
\begin{align}
\mathbf{105}\times \mathbf{20'} &= \mathbf{20'} + \mathbf{105} + \mathbf{175} + \mathbf{336} + \mathbf{729} +
\mathbf{735} \nonumber\\
\mathbf{84}\times \mathbf{20'} &= \mathbf{20'} + \mathbf{84} + \underbrace{\mathbf{70}}_{\mathbf{35} +
\mathbf{\ovr{35}}} + \underbrace{\mathbf{90}}_{\mathbf{45} + \mathbf{\ovr{45}}} + \mathbf{175} +
\underbrace{\mathbf{512}}_{\mathbf{256} + \mathbf{\ovr{256}}} +  \mathbf{729} \;, \label{eqn:prodotti di rappresentazioni}
\end{align}

We shall analyze only the leading term in the OPE expansion parameter, that can be interpreted as contributions coming from Lorentz scalars with bare dimension
$\Delta^{(0)}=4$.
In the representations written above, with these quantum numbers there are the two $\frac{1}{2}$ BPS operator in the $\mathbf{105}$
(that do not appear beyond tree level), and six operators belonging to the $\mathbf{20'}$ representation,
two of them, $\mathcal K_\mathbf{20'}^\pm$, belonging to the Konishi supermultiplet, while the other four are superprimaries.
These are the protected operator $\mathcal{D}_\mathbf{20'}$, and the three superprimary operators $\mathcal{O}_i$.
We have also two operators in the $\mathbf{84}$ with $\Delta^{(0)}=4$, namely $\mathcal D_\mathbf{84}$
and $\mathcal K_\mathbf{84}$, another superdescendant of the Konishi operator.
These operators in principle can appear only in the function $G_{dd}$,
but it is easy to check that $\mathcal K_\mathbf{84}$ does not contribute while $\mathcal D_\mathbf{84}$ gives only tree level contributions.

So, at quantum level we have to deal with the six operators in the $\mathbf{20'}$ representation.
Actually only the three $\mathcal O_i$ pass in the expansions of these functions.
In fact, as we told before, the $\mathcal{D}_\mathbf{20'}$ operator is protected and
its trilinear couplings with the external operators do not renormalize at order $g^2$.
The two Konishi superdescendants do not appear in this channel, either at tree level or at leading logarithms,
due to the fact that they are bilinear in fermions at leading order.
One can verify that this is true also for the order $g$ trilinear coupling\footnote{Being bilinear
in fermions these operators have odd $g$ expansion in the trilinear coupling with the operators in the external points.}.

From these four point functions we can solve the OPE equations (\ref{eqn:sommatreelevel}-\ref{eqn:somma2loop})
for the three operators $\mathcal O_i$. The results of this analysis are reported in \cite{ratio}.

\section{Comments and Outlook}
We developed a technique which allows to calculate (at order $g^4$)
the set of four point functions of protected operators defined by equation \eqref{eqn:prima definizione}
by simply performing some color contractions.
Particular flavour representatives of the operators are considered.
From these correlators, via an OPE analisys, one can extract the
anomalous dimensions and the trilinear couplings of the operators that
appear in the intermediate channel.
Our considerations can be extended to a more general class of protected operators,
in particular to those belonging to general BPS multiplets.

We presented two explicit examples in which this method was used to extract information about the operators in the intermediate channel.
We calculated the anomalous dimensions of the
Lorentz scalar of bare dimension $\Delta^{(0)}=3$ in the
representation $[0,1,0]=\mathbf 6$,
and of the three operators $\mathcal O_i$ in the
representation $[0,2,0]=\mathbf {20'}$ of bare dimension
$\Delta^{(0)}=4$,
together with their trilinear couplings with the external operators.
Moreover, for $\frac{1}{2}$ BPS operators the four point functions
\eqref{eqn:prima definizione} are calculated also for $n=5$ and $n=6$.

One may ask how restrictive is our particular flavour projection.
It would be interesting to study this especially in the case of only $\frac{1}{2}$ BPS operators.
In other terms one has to find how many independent space time
structures at order $g^4$ has this class of correlation functions.
A preliminary analysis suggests that there are at least two independent structures.
Hence in order to take them into account we have to consider at least two
suitable choices of the flavour representatives.
\section*{Acknowledgements}
We thank Ya. S. Stanev for enlightening discussions and comments on the manuscript.
This work was supported in part by INFN, by the MIUR-COFIN contract
2003-023852,
by the EU contracts MRTN-CT-2004-503369 and MRTN-CT-2004-512194, by the INTAS
contract 03-51-6346, and by the NATO grant PST.CLG.978785.
\clearpage
\section*{Appendix}
\subsection*{Notation and conventions}

The field content of ${\cal N}$=4 SYM~\cite{Gliozzi:1976qd} comprises a
vector, $A_{\mu}$, four Weyl spinors, $\psi^{A}$ ($A$=1,2,3,4), and
six real scalars, $\varphi^{i}$ ($i$=1,2,\ldots,6),
all in the adjoint representation of the gauge group $SU(N)$.
In the ${\cal N}$=1 formalism the fundamental fields can be arranged
into a vector superfield, $V$, and three chiral superfields,
$\Phi^{I}$ ($I$=1,2,3). The six real scalars, $\varphi^{i}$, are
combined into three complex fields, namely
\begin{equation}
    \phi^{I} = \frac{1}{\sqrt{2}} \left( \varphi^{I}+i\varphi^{I+3}
    \right) \,,  \qquad \phi^{\dagger}_{I} = \frac{1}{\sqrt{2}} \left(
    \varphi^{I}-i\varphi^{I+3} \right)\,,
    \label{defphi}
\end{equation}
that are the lowest components of the superfields $\Phi^{I}$ and
$\Phi^{\dagger}_{I}$, respectively.
Three of the Weyl fermions, $\psi^{I}$, are the spinors of the
chiral multiplets. The fourth spinor, $\lambda = \psi^{4}$, together
with the vector, $A_{\mu}$, form the vector multiplet.
In this formulation only a $SU(3)\times U(1)$ subgroup of the
original $SU(4)$ R-symmetry group is manifest. $\Phi^{I}$ and
$\Phi^{\dagger}_{I}$ transform in the representations {\bf 3} and
${\overline {\bf 3}}$ of the global $SU(3)$ ``flavour'' symmetry,
while $V$ is a singlet.
We can choose the normalization of  the axial $U(1)$ R-symmetry
such that the three complex scalars $\phi^{I}$ have charge $+1$,
the vector $A_{\mu}$ is neutral,
the gaugino  $\lambda$ has charge $+3/2$ and
the spinors of the chiral multiplets $\psi^{I}$ have charge $-1/2$.
The $I$ flavour index can be raised and lowered with $\delta_{IJ}$,
so we make no difference between $\phi^{I}$ and $\phi_{I}$.

The generators of the gauge group are taken in the fundamental
representation with Dynkin index $d(N)=\frac{1}{2}$.
To perform the contractions we use also the fusion and the fission rules
\begin{align}
\text{tr}\left (T_a A\right)\,\text{tr}\left (T_a B\right) &=
\frac{1}{2} \left( \text{tr}\left(A B \right) -\frac{1}{N} \text{tr} \left(A\right)\, \text{tr} \left(B\right)  \right)\;, \\
\text{tr}\left (T_a A \;T_a B\right) &=
\frac{1}{2} \left( \text{tr} \left(A\right) \,\text{tr} \left(B\right)   -\frac{1}{N} \text{tr}\left(A B \right)\right)\;.
\end{align}

We shall use Fermi-Feynman gauge as it
makes the correction to the propagators of the fundamental
superfields vanish at order $g^{2}$~\cite{Grisaru:1980nk}.
Actually a stronger result holds, namely the anomalous
dimension of the fundamental fields vanish also at $g^4$.
The terms in the action relevant for the calculation of the Green function we are interested in are
\bea
  S_{\rm 0} & = & \int d x \int d^{4}\theta \left( \Phi^{a
  \dagger}_{I}\Phi^{I}_{a} -  V^{a}  \Box V_{a} \right) \label{S0}  \\
  S_{\rm cc} & = & \int d x \int d^{4}\theta
  \, g \frac{{-}{i}{\sqrt{2}}}{{3!}} \left( \varepsilon_{IJK}
  f^{abc}\Phi^{I}_{a}\Phi^{J}_{b}
  \Phi^{K}_{c}\delta^{2}(\overline{\theta}) -
  \varepsilon^{IJK}f_{abc}\Phi^{{a}{\dagger}}_{I}
  \Phi^{{b}{\dagger}}_{J}\Phi^{{c}{\dagger}}_{K}\delta^{2}({\theta})
   \right)\hspace{0.8cm} \label{Scc}  \\
   S_{\rm cv} & = & \int d x \int d^{4}\theta \left(
   2igf_{lmn}\Phi^{{l}{\dagger}}_{I}V^{m}\Phi^{{I}{n}} -
   2g^{2} f_{lmn}f^{n}_{\,\,\,\,qr}\Phi^{{l}
   {\dagger}}_{I}V^{m}V^{q}\Phi^{{I}{r}}
   \right) \label{Scv}   \\
   S_{\rm vv} & = & \int d x\int d^{4}\theta \left(\frac{i{g}}{4}f_{abc}
   \left[ \overline{D}^{2}D^{\alpha}V^{a} \right]V^{b} \left( D_{\alpha}V^{c}
   \right)  \right)\label{Svv}
\eea where $f_{abc}$ are the structure constant of the gauge group $SU(N)$. Due to the fact that all superfields are
massless their free propagators have an equally simple form both in momentum and in coordinate space and we choose to
work in the latter which is more suitable for the study of conformal field theories. The propagators of the chiral and
the vector field are
\bea
 \langle
 \Phi^{I}_{a}(x_{i},\theta_{i},\overline{\theta}_{i})
 \Phi^{{b}{\dagger}}_{J}(x_{j},\theta_{j},\overline{\theta}_{j})\rangle
 & = & \frac{{\delta_{a}^{b}}{\delta_{J}^{I}}}{4\pi^{2}}
 e^{(\xi_{ii}^{\mu}+\xi_{jj}^{\mu}
 -2\xi_{ij}^{\mu})\partial_{\mu}^{i}}\frac{1}{x^{2}_{ij}}\,,
 \label{propchirfree}  \\
 \langle
V^{a}(x_{i},\theta_{i},\overline{\theta}_{i})V_{b}(x_{j},\theta_{j},
\overline{\theta}_{j}) \rangle & = & -
\frac{\delta^{a}_{b}}{8\pi^{2}}
\frac{\delta^{4}(\theta_{ij})}{x_{ij}^{2}} \,,\label{propvetfree}
\eea
where
\bea
  x_{ij} = x_{i}-x_{j}\,, \qquad  \theta_{ij} = \theta_{i}-\theta_{j}\,, \qquad
  \xi^{\mu}_{ij}=\theta^{\a}_{i}\s^{\, \mu}_{\a\da}
  \overline{\theta}^{\da}_{j}\,, \qquad
  \partial_{\,\mu}^{\,i} = \frac{\partial}{\partial x_{i}^{\,\mu}}\,.
  \nonumber
\eea

\subsection*{Computation of the relevant integrals}
The whole result, in the limit $x_1 \rightarrow \infty$ (see \eqref{eqn:definizione di limite}) can be expressed in
terms of two functions
\begin{gather}
g(i,j,k)  = \int {dx_5 \over x_{i5}^2 x_{j5}^2 x_{k5}^2}
 = \, \frac{\pi^{2}}{x_{jk}^{2}}
\, B \left( \frac{x_{ij}^{2}}{x_{jk}^{2}} \, ,
\frac{x_{ik}^{2}}{x_{jk}^{2}} \right)\,,
\label{defg}  \\
f(i;j,k)  = \int {dx_5 dx_6 \over x_{i5}^2 x_{j5}^2 x_{i6}^2 x_{k6}^2 x_{56}^2}
=\, \frac{\pi^{4}}{x_{jk}^{2}} \, \Phi^{(2)} \left(
\frac{x_{ij}^{2}}{x_{jk}^{2}} \, , \frac{x_{ik}^{2}}{x_{jk}^{2}} \right)\;,
\label{deff}
\end{gather}
where 
the two functions $B$ and $\Phi^{(2)}$ can be written as
\cite{Usyukina:1992jd,Bianchi:1999ge,Eden:1998hh}
\begin{gather}
\begin{split}
B(r,s) = &
{1 \over \sqrt{p}} \left \{ \log (r)\log (s)  -
\left [\log \left({r+s-1 -\sqrt {p} \over 2}\right)\right ]^{2}
\right. \\&\left.
-2 {\rm Li}_2 \left({2 \over 1+r-s+\sqrt {p}}\right ) -
2 {\rm Li}_2 \left({2 \over 1-r+s+\sqrt {p}}\right )\right \}\;,
\label{Brsf}
\end{split}
\\
\begin{split}
\Phi^{(2)}(r,s)  = & \frac{1}{\sqrt{p}}
\left\{
 6\left(\, {\rm Li}_{4}(-\r r)+{\rm Li}_{4}(-\r s)\,\right)+
 3\log(\frac{s}{r})\left( {\rm Li}_{3}(-\r r)-{\rm Li_{3}}(-\r s) \right)\,+
 \right.  \\
  & +  \left.
 \frac{1}{2} \log^{2}\left(\frac{s}{r}\right)\left({\rm Li}_{2}(-\r r)+{\rm Li}_{2}(-\r s) \right)+
 \frac{1}{4} \log^{2}(\r r) \log^{2}(\r s)\,+
 \right. \label{Phi2rsf} \\
  & +   \left.
 \frac{\pi^{2}}{2} \log(\r r)\log(\r s)+
 \frac{\pi^2}{12} \log^{2}\left( \frac{s}{r} \right)+
 \frac{7\pi^4}{60}
 \right\}\;,
\end{split}
\end{gather}
with
\begin{align}
p &= 1 + r^{2} + s^{2} - 2r - 2s - 2rs\;, &
\rho&=\frac{2}{1-r-s+\sqrt{p}}\,, &
{\rm Li}_{N}(z) \, &\stackrel{\tiny |z| < 1}{\equiv} \, \sum_{k=1}^{\infty} \, \frac{z^{k}}{k^{\,N}}\;.
\end{align}

Both $B(r,s)$ and $\Phi^{(2)}(r,s)$ are symmetric in $r$ and $s$. Moreover they have singularities in $(r=0,s=1)$,
$(r=1,s=0)$ and $(r=\infty,s=\infty)$. We will write now the expansion of these functions near these singularities,
\begin{align}
B\left(\frac{\eps^2}{x^2},\left(1 + \frac{\eps}{x}\right)^2\right)  &=  2-\log\left(\frac{\eps^2}{x^2}\right)
- \frac{\eps \cdot x}{x^2}\left(1-\log\left(\frac{\eps^2}{x^2}\right)\right)
- \frac{\eps^2}{x^2}\left(\frac{2}{9}-\frac{1}{3}\log\left(\frac{\eps^2}{x^2}\right)\right)\nonumber\\
&+\frac{(\eps \cdot x)^2}{x^4}\left(\frac{8}{9}-\frac{4}{3}\log\left(\frac{\eps^2}{x^2}\right)\right)
+\frac{\eps^2\,\eps \cdot x }{x^4}\left(\frac{1}{2}-\log\left(\frac{\eps^2}{x^2}\right)\right)\\
&-\frac{\eps^2 (\eps \cdot x)^2}{x^6}\left(\frac{24}{25}-\frac{12}{5}\log\left(\frac{\eps^2}{x^2}\right)\right)+\cdots
\nonumber\;.
\end{align}
We do not need the expansion of $B(r,s)$ near the point $(r=\infty,s=\infty)$ because of the relation
\be
B(r,s)= \frac{1}{r} B\left(\frac{1}{r},\frac{s}{r}\right) = \frac{1}{s}B\left(\frac{1}{s},\frac{r}{s}\right)\;.
\ee
For the function $\Phi^{(2)}(r,s)$ we have
\begin{align}
\Phi^{(2)}\left(\frac{\eps^2}{x^2},\left(1 + \frac{\eps}{x}\right)^2\right)  &=
6 - 3\log\left(\frac{\eps^2}{x^2}\right) +\frac{1}{2}\log^2\left(\frac{\eps^2}{x^2}\right)\nonumber\\
&-\frac{1}{4}\frac{\eps \cdot x}{x^2}
\left(21 - 13\log\left(\frac{\eps^2}{x^2}\right) +3\log^2\left(\frac{\eps^2}{x^2}\right)\right)\nonumber\\
&-\frac{1}{36}\frac{\eps^2}{x^2}
\left(\frac{251}{3} - 49\log\left(\frac{\eps^2}{x^2}\right) + 11 \log^2\left(\frac{\eps^2}{x^2}\right)\right)\\
&+\frac{1}{9}\frac{(\eps \cdot x)^2}{x^4}\left(\frac{367}{6}-40\log\left(\frac{\eps^2}{x^2}\right)+11\log^2\left(\frac{\eps^2}{x^2}\right)\right)
+ \cdots
\nonumber\\
\Phi^{(2)}\left(\frac{1}{\left(1 + \frac{x}{\eps}\right)^{2}},\frac{1}{\left(1 + \frac{\eps}{x}\right)^{2}}\right)  &=
6 - 3\log\left(\frac{\eps^2}{x^2}\right) +\frac{1}{2}\log^2\left(\frac{\eps^2}{x^2}\right)\nonumber\\
&+\frac{1}{4}\frac{\eps \cdot x}{x^2}
\left(45 - 21\log\left(\frac{\eps^2}{x^2}\right) + 3\log^2\left(\frac{\eps^2}{x^2}\right)\right)\nonumber\\
&+\frac{1}{9}\frac{\eps^2}{x^2}
\left(\frac{160}{3} - 26\log\left(\frac{\eps^2}{x^2}\right) + 4 \log^2\left(\frac{\eps^2}{x^2}\right)\right)\\
&-\frac{1}{18}{\frac {(\eps \cdot x)^{2}}{{x}^{4}}}\left({\frac {65}{3}}\,
-19\log\left(\frac{\eps^2}{x^2}\right)+5{\log^2\left(\frac{\eps^2}{x^2}\right)}\right)\nonumber
\;.
\end{align}

\subsection*{Feynman Diagrams}
Here we list the Feynman diagrams used in our calculation \cite{Bianchi:2000hn}. The results indicated are all in the
limit $x_1 \rightarrow \infty$, after multiplication by a factor of $x_1^4$ for all the diagrams except $A_5$ and
$B_{22}$ -- $B_{27}$, that are multiplicated by $x_1^6 $. We indicate here also equalities between the diagrams that
are valid only in this limit.
The complete $x_1$ dependence of a correlation function (with well defined conformal properties) can be recovered
after the substitutions $x_{23}^2 \rightarrow x_{23}^2x_{14}^2$, $x_{24}^2 \rightarrow x_{24}^2x_{13}^2$, $x_{34}^2 \rightarrow x_{34}^2x_{12}^2$.
These results are grouped depending of the number of vector lines in it. Diagrams of type
$A$ are without vector lines, the type $B$ contain one vector line and the type $C$ two.

\begin{align}
A_1 & =  \frac{1}{(4\pi^2)^8}\,\frac{1}{x_{34}^{2}}\,f(4;2,3)\,, & A_2 &
=  \frac{1}{(4\pi^2)^8}\,\frac{1}{x_{34}^{2}}\,f(2;3,4)\,,
& A_4 &=  \frac{1}{(4\pi^2)^8}\,\left[g(2,3,4)\right]^2 \,,\notag \\
B_4 & =-\frac{1}{2}\frac{1}{(4\pi^2)^8}\,\frac{1}{x_{34}^{2}}\,f(3;2,4)\,,& &&
A_5 & =  \frac{1}{(4\pi^2)^9}\,\frac{1}{x_{34}^{2}}\,f(3;2,4)\,,
\end{align}
\begin{align}
B_1 &=B_9 =B_{24}=B_{27} = 0\,, & B_8&=B_4\,, & A_3&=A_2\,,
& B_7=B_{12}&=B_{14}=B_{16}=-\frac{1}{2} A_1\,, \nonumber
\\ B_{21}&=B_4-\frac{1}{4} (1+ \frac{s}{r} -\frac{1}{r}) A_4\,, &
C_6&=-\frac{1}{4}\frac{s}{r}A_4\,, & C_7&=-\frac{1}{4}A_1\,, & B_{10}&=B_{11}= -\frac{1}{2} A_2\,.
\end{align}
Here there are other relations, valid in the limit described above, that make some diagrams sums to zero:
\begin{align}
B_5&=B_{20}=B_2\,, & B_6&=B_{13}=B_{17}=B_{19}=B_3\,, & B_{18}&=B_{15}\,, &B_{23}&=B_{25}\,,\\
C_1&=C_3\,, & C_5&=-C_9=C_4\,, &C_8&=-C_2\,, & B_{27}&=B_{26}\,.
\end{align}

\begin{center}
\begin{table}[!h]

\caption{Diagrams without vector lines.}
\end{table}
\begin{table}[!h]
%
\caption{Diagrams with two vector lines.}
\end{table}
\newpage
\begin{table}[!h]
%
\caption{Diagrams with one vector line}
\end{table}


\end{center}

\clearpage
\subsection*{Four point functions of higher dimensional operators}
Here we indicate the correlators $G_n$, $n=5,6$ with only $\frac{1}{2}$ BPS operators in the external points.
Their complete form can be reconstructed from equations (\ref{eqn:espansione in potenze di Gn},
\ref{eqn:contrazioni indipendenti}-\ref{eqn:risultato di G0}).
The $SU(4)$ representations under consideration are $[0,5,0]=\mathbf{196}$ and $[0,6,0]=\mathbf{336}$.
These operators are normalized so that their two point function is
\be
\langle {\mathcal{O}}_{[0,n,0]}^{\Delta=n}(x) {\mathcal{O}}_{[0,n,0]}^{\Delta=n}(0)\rangle = \frac{1}{(4 \pi^2)^n} \frac{M_{\mathcal O}}{x^{2n}}\;.
\ee
We indicate schematically (following the notation of eq. \eqref{eqn:representatives}, with complete symmetrization of the color indices) the definition of the operators,
and the normalization of their two point functions:
\begin{itemize}
\item $n=5$
\begin{align}
\mathcal Q_\mathbf{196} &\equiv \text{tr} \left( \phi^5\right) & M_{\mathcal Q_\mathbf{196}} &= \frac{(N^2-1)(N^2-4)(N^4+24)}{8 N^3}\;, \nonumber\\
\mathcal D_\mathbf{196} &\equiv \text{tr} \left( \phi^3\right) \text{tr} \left( \phi^2\right) & M_{\mathcal D_\mathbf{196}} &= \frac{15 (N^2-1)(N^2-4)(N^2+5)}{N}\;;
\end{align}
\begin{table}[hbtp]
\begin{center}
\begin{tabular}{||c||c|c||}
\hline
\hline
Function & $Y\cdot \delta^2$ & $Y \cdot H_1$  \\
\hline
\rule{0pt}{4ex}
$G_{\cal D_\mathbf{196}D_\mathbf{196}} $&$\frac{(N^2-1)(N^2-4)(N^2+5)^2}{4 N}$   & $\frac{5 (N^2-1)(N^2-4) (N^2+5)}{8}$                                  \\
\rule{0pt}{4ex}
$G_{\cal D_\mathbf{196}Q_\mathbf{196}}$ &$\frac{(N^2-1)(N^2-4)(N^2-2)(N^2+5)}{8 N^2}$   & $\frac{5 (N^2-1)(N^2-4)(N^2-2)}{16 N} $ \\
\rule{0pt}{4ex}
$G_{\cal Q_\mathbf{196}Q_\mathbf{196}}$ &$\frac{ (N^2-1)(N^2-4)(N^2-2)^2}{24 N^3}$   & $\frac{(N^2-1)(N^2-4)(N^4 +24)}{192 N^2} $ \\
\hline
\hline
Function  & $Y\cdot H_2$ & $Z \cdot L_1$\\
\hline
\rule{0pt}{4ex}
$G_{\cal D_\mathbf{196}D_\mathbf{196}} $  &$ \frac{7 N (N^2-1)(N^2-4) (N^2+17)}{16} $ & $-\frac{N (N^2-1)(N^2-4)(N^2-23)}{16} $ \\
\rule{0pt}{4ex}
$G_{\cal D_\mathbf{196}Q_\mathbf{196}}$  &$ \frac{(N^2-1)(N^2-4)(9 N^2+10)}{32} $ &$ -\frac{(N^2-1)(N^2-4) (N^2-30)}{96} $ \\
\rule{0pt}{4ex}
$G_{\cal Q_\mathbf{196}Q_\mathbf{196}}$ &$ \frac{(N^2-1)(N^2-4)(N^4+24 N^2 -24)}{384 N}$ &$-\frac{(N^2-1)(N^2-4)(N^4 -24 N^2 +72)}{1152 N}$\\
\hline
\hline
\end{tabular}
\end{center}
\caption{Color contractions for functions of type $G_5$}\label{Color contraction for functions of $G_5$ type}
\end{table}
\item $n=6$
\begin{align}
\mathcal O^{6} &\equiv \text{tr} \left( \phi^6\right) & M_{\mathcal O^{6}} &= \frac{45 (N^2-1)(N^8+6 N^6 -60 N^4 +600)}{8 N^4}\;,\nonumber \\
\mathcal O^{33} &\equiv \text{tr} \left( \phi^3\right)^2  &
M_{\mathcal O^{33}} &= \frac{135 (N^2-1)(N^2-4)^2(N^2+8)}{8 N^2}\;,\nonumber\\
\mathcal O^{42} &\equiv \text{tr} \left( \phi^4\right) \text{tr} \left( \phi^2\right) &
M_{\mathcal O^{42}} &= \frac{30 (N^2-1)(N^6 +17 N^4 -72 N^2 +162)}{N^2}\;,\\
\mathcal O^{222} &\equiv \text{tr} \left( \phi^2\right)^3 &
M_{\mathcal O^{222}} &= 2880(N^4-1)(N^2+3)\;;\nonumber
\end{align}
\begin{table}[hbtp]
\begin{center}
\begin{tabular}{||c||c|c||}
\hline
\hline
Function & $Y\cdot \delta^2$ & $Y \cdot H_1$  \\
\hline
\rule{0pt}{4ex}
$G_{\mathcal O^{6}\mathcal O^{6}} $&$\frac{3(N^2-1)(4 N^8 -26 N^6 +155 N^4 -450 N^2 +450)}{32 N^4}$   & $\frac{3 (N^2-1)(N^8+6 N^6 -60 N^4 +600)}{128 N^3}$ \\
\rule{0pt}{4ex}
$G_{\mathcal O^{6}\mathcal O^{33}} $&$\frac{9(N^2-1)(N^2-4)(2 N^4 -15 N^2 +30)}{32 N^3}$   & $\frac{9 (N^2-1)(N^2-4)(3 N^4 -20 N^2 +40)}{128 N^2}$ \\
\rule{0pt}{4ex}
$G_{\mathcal O^{6}\mathcal O^{42}} $&$\frac{(N^2-1)(2 N^8 +41 N^6 -211 N^4 + 630 N^2 - 810)}{16 N^3}$   & $\frac{3 (N^2-1)(2 N^6 -13 N^4 +60 N^2 - 90)}{16 N^2}$ \\
\rule{0pt}{4ex}
$G_{\mathcal O^{6}\mathcal O^{222}} $&$\frac{15 (N^2-1)(N^2+3)(N^4 -3 N^2 +3)}{2 N^2}$   & $\frac{15 (N^2-1)(N^4-3 N^2+3)}{2 N}$ \\
\rule{0pt}{4ex}
$G_{\mathcal O^{33}\mathcal O^{33}} $&$\frac{27(N^2-1)(N^2-4)(N^2 -8)}{32 N^2}$   & $\frac{9 (N^2-1)(N^2-4)^2(N^2 + 8)}{128 N}$ \\
\rule{0pt}{4ex}
$G_{\mathcal O^{33}\mathcal O^{42}} $&$\frac{3(N^2-1)(N^2-4)(N^4 +9 N^2 - 54)}{16 N^2}$   & $\frac{9 (N^2-1)(N^2-4)(N^2 -6)}{16 N}$ \\
\rule{0pt}{4ex}
$G_{\mathcal O^{33}\mathcal O^{222}} $&$\frac{27(N^2-1)(N^2-4)(N^2+3)}{6 N}$   & $\frac{9 (N^2-1)(N^2-4)}{2}$ \\
\rule{0pt}{4ex}
$G_{\mathcal O^{42}\mathcal O^{42}} $&$\frac{(N^2-1)(N^8 +72 N^6 + 47 N^4- 666 N^2 +1458)}{24 N^2}$   & $\frac{(N^2-1)(N^6 +17 N^4 -72 N^2 +162)}{8 N}$ \\
\rule{0pt}{4ex}
$G_{\mathcal O^{42}\mathcal O^{222}} $&$\frac{3(N^2-1)(N^2+3)^2(2N^2-3)}{N}$   & $\scriptstyle 3(N^2-1)(N^2+3)(2N^2-3)$ \\
\rule{0pt}{4ex}
$G_{\mathcal O^{222}\mathcal O^{222}} $&$\scriptstyle 12(N^4-1)(N^2+3)^2$   & $\scriptstyle 12N (N^4-1)(N^2+3)$ \\
\hline
\hline
Function  & $Y\cdot H_2$ & $Z \cdot L_1$\\
\hline
\rule{0pt}{4ex}
$G_{\mathcal O^{6}O^{6}} $  &$ \frac{3(N^2-1)(N^8+47 N^6 -324 N^4 +1200 N^2 -1200)}{256 N^2} $ &
$-\frac{3(N^2-1)(N^8 -35 N^6 +204 N^4 -1200 N^2 +2400)}{1024 N^2}$ \\
\rule{0pt}{4ex}
$G_{\mathcal O^{6}\mathcal O^{33}} $&$\frac{9(N^2-1)(N^2-4)( N^4 -20)}{64 N}$   & $-\frac{9 (N^2-1)(N^2-4)(N^4-20 N^2 +80)}{512 N}$ \\
\rule{0pt}{4ex}
$G_{\mathcal O^{6}\mathcal O^{42}} $&$\frac{5(N^2-1)(N^6+ 5 N^4 -18 N^2 - 54)}{16 N}$   & $-\frac{ (N^2-1)(N^6 - 64 N^4+270 N^2 -540)}{64 N}$ \\
\rule{0pt}{4ex}
$G_{\mathcal O^{6}\mathcal O^{222}} $&$\frac{3(N^2-1)(7 N^4 +2 N^2 - 15)}{2}$   & $-\frac{3 (N^2-1)(N^2+10)(2 N^2 -3)}{8}$ \\
\rule{0pt}{4ex}
$G_{\mathcal O^{33}\mathcal O^{33}} $&$\frac{9(N^2-1)(N^2-4)( N^4+19 N^2 -80)}{256}$   & $-\frac{9 (N^2-1)(N^2-4)(N^4-11 N^2 +16)}{1024}$ \\
\rule{0pt}{4ex}
$G_{\mathcal O^{33}\mathcal O^{42}} $&$\frac{3(N^2-1)(N^2-4)( 2 N^2 -9)}{8}$   & $\frac{3N^2(N^2-1)(N^2-4)}{64}$ \\
\rule{0pt}{4ex}
$G_{\mathcal O^{33}\mathcal O^{222}} $&$\scriptstyle 9 N (N^2-1)(N^2-4)$   & $\frac{9 N(N^2-1)(N^2-4)}{8}$ \\
\rule{0pt}{4ex}
$G_{\mathcal O^{42}\mathcal O^{42}} $&$\frac{(N^2-1)(N^6+53 N^4-39 N^2 +189)}{12}$   & $-\frac{(N^2-1)(N^2+3)(N^4-58 N^2 +36)}{96}$ \\
\rule{0pt}{4ex}
$G_{\mathcal O^{42}\mathcal O^{222}} $&$\scriptstyle N (N^2-1)(7 N^4 + 62 N^2 - 45)$   & $\frac{N(N^2-1)(N^4+53 N^2-18)}{4}$ \\
\rule{0pt}{4ex}
$G_{\mathcal O^{222}\mathcal O^{222}} $&$\scriptstyle 12 N^2(N^2-1)(N^2+3)(N^2+11)$   & $\scriptstyle 30 N^2 (N^2-1)(N^2+3)$ \\
\hline
\hline
\end{tabular}
\end{center}
\caption{Color contractions for functions of type $G_6$}\label{Color contraction for functions of $G_6$ type}
\end{table}
\end{itemize}

\end{fmffile}
\clearpage
\providecommand{\href}[2]{#2}\begingroup\raggedright\endgroup

\end{document}